# Scrutinizing the Debye plasma model: Rydberg excitons unravel the properties of low-density plasmas in semiconductors

Heinrich Stolz[1], Dirk Semkat[2], Rico Schwartz[1], Julian Heckötter[3], Marc Aßmann[3], Wolf-Dietrich Kraeft[1], Holger Fehske[2], and Manfred Bayer[3]

[1]Institut für Physik, Universität Rostock, Albert-Einstein-Str. 23, 18059 Rostock, Germany,
[2]Institut für Physik, Ernst-Moritz-Arndt-Universität Greifswald, Felix-Hausdorff-Str. 6, 17489 Greifswald, Germany,
[3]Experimentelle Physik 2, Technische Universität Dortmund, 44221 Dortmund, Germany.

*For low-density plasmas, the classical limit described by the Debye-Hückel theory is still considered as an appropriate description even though a clear experimental proof of this paradigm is lacking due to the problems in determining the plasma-induced shift of single-particle energies in atomic systems. We show that Rydberg excitons in states with a high principal quantum number are highly sensitive probes for their surrounding making it possible to unravel accurately the basic properties of low-density non-degenerate electron-hole plasmas. To this end, we accurately measure the parameters of Rydberg excitons such as energies and linewidths in absorption spectra of bulk cuprous oxide crystals in which a tailored electron-hole plasma has been generated optically. Since from the absorption spectra exciton energies, as well as the shift of the single-particle energies given by the band edge, can be directly derived, the measurements allow us to determine the plasma density and temperature independently, which has been a notoriously hard problem in semiconductor physics. Our analysis shows unambiguously that the impact of the plasma cannot be described by the classical Debye model, but requires a quantum many-body theory, not only for the semiconductor plasma investigated here, but in general. Furthermore, it reveals a new exciton scattering mechanism with coupled plasmon-phonon modes becoming important even at very low plasma densities.*

I. Introduction

Plasmas, a notion coined by Langmuir [1] have been central to the physics of matter in the last 100 years. Since plasmas, consisting in general of electrically charged and neutral particles, represent the most common state of matter in the universe, a profound understanding of their properties is important in many fields ranging from the evolution of the universe, star formation, and fusion to a vast number of technological and even medical applications (for recent reviews see [2,3,4]).

An important question in this field is the formation and nature of bound-states in a plasma, as their observation allows investigating the properties of a plasma from within [5]. The retroactive effects of plasmas on the bound states are three-fold: (i) shift of the binding energies, (ii) broadening of the absorption and emission lines, and (iii) lowering of the two-particle scattering continuum edge. The issues (i) and (iii) together lead to a lowering of the ionization energies, and, therefore, finally to the vanishing of the bound states. This behavior is usually referred to as the Mott effect [6,7].

The fundamental interaction law between the charged particles in a plasma has been derived already by Debye and Hückel in 1923 [8], in the framework of classical statistical thermodynamics, leading to the famous screened Debye potential

$$V_D(r) = \frac{1}{4\pi\varepsilon_0\varepsilon_r} \frac{e^{-\kappa_D r}}{r} \;, \quad (1)$$

with the inverse Debye screening length

$$\kappa_D = \left(\frac{\rho e^2}{\varepsilon_0\varepsilon_r k_B T}\right)^{\frac{1}{2}} . \quad (2)$$

Here $\rho$ denotes the particle density, $T$ the plasma temperature, and $\varepsilon_r$ the relative dielectric constant of the medium. All other symbols have their usual meaning.

This screened interaction leads to a shift of the single-particle energies which amounts to

$$\Delta E_{sp} = -\frac{1}{2}\frac{e^2}{4\pi\varepsilon_0\varepsilon_r}\kappa_D \;. \quad (3)$$

However, it wasn't until 1956 that Ecker and Weizel solved the Schrödinger equation for the two-particle problem with interaction (1) modified by the subtraction of the shift of the single-particle energies (3) and found for small $\kappa_D$ for the shift of energies of the bound-states $\Delta E_X(n) \propto n^2 Ry_X \kappa_D^2$ with the principal quantum number $n$. In this model, they were able to explain many open questions of plasma physics like the vanishing of bound states in dense plasmas [9]. Since then it has been a paradigm of plasma physics that non-degenerate plasmas can approximately be described by the Debye-Hückel-Ecker-Weizel theory, in short, the "Debye" model [6,7]. For these plasmas, the product of density $\rho$ and thermal de Broglie length $\Lambda_{th} = \sqrt{(2\pi\hbar^2)/(\mu k_B T)}$ to the power of the dimension, $\rho\Lambda_{th}^d$ (here $\mu$ is the reduced mass of the plasma constituents) is much smaller than unity, and the distribution function of the plasma constituents is the classical Boltzmann distribution.

While many areas of plasma physics are founded on concepts of classical physics [2-4], an in-depth description of plasmas can only be achieved within a full quantum many-body theory. This was first achieved by various authors in the 1970s [10-14], for reviews see [6,7,15]. Currently, this subject is discussed under the topic of "ionization-potential depression" (IPD), see, e.g. Refs. 16-20. In this context, the question arises whether the "classical limit" of these many-body theories is still given by the Debye plasma model. Unfortunately, the validity of the Debye description for atomic plasmas is extremely difficult to test, as here the determination of the shift of the absolute single-particle energies (Eq. (3)) and bound-state energies is impossible, only relative values can be measured as IPD [3,21,22] or as an energy difference between bound-states in spectroscopic measurements [3].

After the early years of plasma physics, it became obvious that not only atomic systems like hydrogen have plasma phases, but most interacting systems. A prominent example is the system of interacting quarks and gluons in the early state of the universe [23,24], but also plasma phases in semiconductors formed from interacting quasi-particles, electron-hole plasmas (EHPs), have gained huge interest [25-32]. For example, the important field of light emission from semiconductors (lasers and LED) is determined by the properties of the EHP produced in these devices. Compared to atomic systems, the main advantage of these EHPs is that they not only can be produced "at-will", e.g. by shining light of photon energy in the continuum, i.e., above the semiconductor band edge, onto the sample, but one can obtain both the absolute position of the single-particle ground state, which per se is identical with the band-edge energy $E_g$ and the absolute bound-state energies, i.e., the excitons, by a simple measurement of the absorption spectrum [26,27]. Indeed, systematic investigations in the 1970s and 1980s and ongoing up to now have led to great insights into the physics of semiconductors [25-32].

So far the investigations of the effects of plasmas on bound-states in semiconductors have been restricted almost exclusively to the study of the lowest bound-state, which is the analog of the 1S state of the hydrogen atom. To obtain observable plasma effects in this case densities far beyond the degeneracy limit are required and a thorough theoretical treatment of these plasmas is only possible within quantum many-particle theories [25,28,29]. To investigate the non-degenerate limit at moderate temperatures would require very low densities, e.g. for GaAs at $T = 10$ K much smaller than $10^{14}\,\text{cm}^{-3}$, which has been below the experimental accessibility so far [26].

The recent observation of Rydberg excitons, i.e., electron-hole bound-states with very high principal quantum numbers up to 30, in cuprous oxide [33-35], opened up the possibility to study the interaction of such weakly bound excitons with Bohr radii of the order of 1 μm with plasmas of low density, where the Debye model is expected to be a good description. Indeed, the first study [37] of Rydberg excitons in the presence of an EHP with densities around $10^{10}$ cm$^{-3}$ showed a quenching of the absorption of P states with high quantum numbers. However, interpreting the experimental results within the Debye model raised several problems, which could not be solved unambiguously. The first one was the observation that even without any additional pumping the area of the P absorption lines decreases with quantum number much faster than expected from exciton theory [38]. Recently, this effect could be explained, at least qualitatively, by the influence of uncompensated residual charged impurities, which are present in all real crystals and produce a reduction of oscillator strength of Rydberg states and an increase of linewidth with high quantum numbers by ionizing these states in their static random electric field [39]. The unique signature of this effect is the occurrence of an Urbach tail, i.e., an absorption band that continues the continuum absorption towards lower energies with an exponential line shape. The second and much more severe one was the discrepancy between the shift of the band edge that could be measured with great accuracy and the lack of the corresponding energy shift of the exciton states, which is predicted in the Debye model [40]. However, the plasma created in this experiment is a rather complex one. It consists of electrons and holes created directly by the pump laser with photon energy above the band edge as well as those created indirectly by the Auger decay [41,42] of excitons, resulting in quite different temperatures of the components. As the temperature cannot be obtained independently from the experiments, but directly influences both band-edge and exciton-energy shifts, the comparison between theory and experiment is very difficult. As a consequence, the important question, how the plasma quenching competes with the simultaneously expected excitonic blockade effect [33,39] could not be decided from these studies.

Recently, an elaborate many-particle theory has been applied to Rydberg excitons interacting with an ultra-cold, low-density EHP [43,44]. The predictions are quite different from those of the classical Debye model. First, the effective band-edge energy shifts with the fourth root of the product of plasma density and temperature of the EHP instead and not with the square root of the ratio of density and temperature (see Eqs. (3) and (2)). Second, the energetic position and oscillator strength of an exciton state remains constant up to small deviations of less than 5% in the binding energy and less than 1% in oscillator strength up to the density where the redshifting band edge crosses the energy of the exciton state. This density marks the Mott point, which is considered as the usual criterion for the Mott effect [6,7]). In contrast, the Debye model predicts not only energetic shifts of the order of the binding energies but also losses of the oscillator strength by more than a factor of two [27]. We note that such differences between the results of a more rigorous many-particle approach and the Debye model were found in earlier theory papers [30,31,45,46], but have neither been validated due to the absence of accurate experimental observations, e.g. of the band edge nor taken seriously.

II. Experiment and data analysis

Therefore, in this paper, we try to avoid the complexity introduced by a multi-component plasma by creating a well-defined EHP which solely contains Auger-created electron-hole pairs. We achieve this in a pump-probe setup (for details see Ref. 36), where one tunable single-frequency laser (power $1\,\mu\text{W}$) is scanned to measure the transmission spectrum of the sample $T(\hbar\omega) = P_{out}/P_{in}$ of the yellow P states, whereby $P_{in}$ and $P_{out}$ are the laser powers before and behind the sample, respectively. The sample with a thickness of about $32\,\mu\text{m}$ has been cut off from a high-quality natural $Cu_2O$ crystal from the Tsumeb mine in Namibia, which has been used in previous studies of Rydberg excitons [33,35,37]. The sample is mounted strain-free [47] and have been immersed into superfluid helium in an optical cryostat providing $T = 1.35$ K to provide optimal cooling. The input and output powers have been obtained by correcting for the reflection losses due to the cryostat quartz windows.

To produce the plasma, we additionally pump with a second laser (power from 0.2 μW to 5 mW) into pure yellow 1S states (in the phonon-assisted indirect absorption, see the blue arrow in Fig. 1) at a temperature of 1.35 K. At this energy, neither free holes nor P excitons are created by the laser. Since the plasma in this case is simple, we expect it to be describable by many-body theory allowing a direct comparison with experiment. To this end, one has to measure accurately the spectroscopic position of both the band edge and of the Rydberg states as a function of the plasma density, which can be varied by changing the pump laser power. Here we have to ensure that the experiments are performed with high enough plasma densities that the influence of residual charged impurities can be neglected, proven by the non-existence of an Urbach tail. On the other hand, the densities have to be small enough to stay in the nondegenerate regime, i.e., $\rho_{eh}\Lambda_{th}^{d} \ll 1$.

A closer look at the experimental transmission spectra (full colored lines in Fig.1) already reveals that we have achieved this goal. At very low pump powers (compare the spectra at 0.2 μW and 2 μW) the spectra show pronounced dips in the transmission due to Rydberg excitons up to a maximum principal quantum number of $n = 23$ [48], where the transmission changes into the smooth continuum response of the scattering states. The energy of the transition point is marked by a red arrow and is considered as the shifted apparent band dge. With increasing pump power the spectra change consistently. The peaks corresponding to the highest principal quantum number $n$ lose strength until they vanish in the red-shifting continuum. We attribute this to the effect of the increasing density of an EHP leading to a redshift of the band edge and a concomitant shift of the Mott point. The dips at low $n$ seem to be unaffected by the plasma effect.

To deduce quantitatively the parameters of the exciton states (energy, oscillator strength, and linewidth) and the band-edge position from the transmission spectra (for an overview see Fig. 1) we use the recently developed highly accurate coherence transfer matrix (CTM) method [46]. The essence of this way of calculating the transmission spectrum of a thin plate of semiconductors is to model the spectral dependence of the complex dielectric function $\varepsilon(\hbar\omega)$ around the yellow band edge by assuming contributions from all relevant optical resonances. Defining appropriate transfer matrices one can take into account the coherence of the measuring laser beam (coherence length several km), a spectrally varying reflection coefficient, possible oxide layers on top of the sample, and scattering due to surface roughness (for details see Ref. 46).

The results of the fit by this method are given by the dotted lines in Fig. 1. The almost quantitative agreement of the calculated and measured transmission spectra demonstrates the high accuracy of this method.

To describe the optical response of $Cu_2O$ in the region around the yellow gap, the following contributions to $\varepsilon(\hbar\omega)$ are essential: (i) the contribution of the P exciton transitions and the corresponding continuum, (ii) the indirect absorption into the yellow and green exciton states [50][49], and (iii) the background contribution from exciton states corresponding to higher bands[51]

At this point, we give only the imaginary part of the dielectric functions for the P states (all other contributions to the dielectric function can be found in Ref. 46):

$$\varepsilon_{i,P}(\hbar\omega) = \sum_{n=2}^{n_{max}} \tilde{f}_n \frac{1}{\pi} \frac{\Gamma_n + 2A_n \cdot (\hbar\omega - E_n)}{(\hbar\omega - E_n)^2 + \Gamma_n^2 \cosh\left(\frac{\hbar\omega - E_n}{3\Gamma_n}\right)^2} \quad . \qquad (4)$$

In Eq. (4), the sum goes over the P states (principal quantum number $n$) and the $\tilde{f}_n$ are constants proportional to the oscillator strengths, $E_n$ are the energies of the P states. The damping $\Gamma_n$ and the asymmetry $A_n$ are assumed to be constant for each P state. The factor $\cosh(\zeta)^2$ with $\zeta = (\hbar\omega - E_n)/3\Gamma_n$ has been introduced to obtain an exponentially damped line shape at energies far from the resonance (Urbach-like tail) [49]. The maximum quantum number $n_{max}$, up to which excitons exist, is determined by the shift of the apparent band edge $\Delta = E_g - E_{g0}$ ($E_{g0} = 2.172049$ eV being the nominal gap [52]), since above this energy the Mott effect appears [6,7,14,15,29]. The P continuum absorption sets in at the band edge with a step-like increase of the absorption due to Sommerfeld enhancement but is broadened by a Lorentzian with damping $\gamma_c$ [46]

$$\varepsilon_c(\hbar\omega) = a_c \left(1 + \frac{\hbar\omega - E_g}{Ry}\right) \left[\frac{1}{2} + \frac{1}{\pi} \arctan\left(\frac{\hbar\omega - E_g}{\gamma_c}\right)\right] , \qquad (5)$$

where $a_c = 1.645 \cdot 10^{-3}$ [49] is a factor describing the strength of the continuum, $E_g$ is the apparent band edge, and $Ry$ is the exciton Rydberg energy. We expect $\gamma_c$ as a two-particle property (damping of electron-hole pairs in scattering states) to be of the order of the exciton damping at the band edge.

The analysis of the interaction of Rydberg excitons with charged impurities [39] shows the existence of an Urbach tail, which continues the continuum to lower energies, being of the general form

$$\varepsilon_{UT}(\hbar\omega) = a_c e^{\frac{-E_g + \hbar\omega}{E_{ur}}} \Phi(E_g - \hbar\omega) \qquad (6)$$

with $E_{ur}$ the decay constant of the tail. To obtain a smooth transition from the continuum to the Urbach tail we broaden it by a Lorentzian with the same width as the continuum (for details see Appendix I).

Clearly, the real part of the dielectric function can be obtained by a Kramers-Kronig transform of the imaginary part [49].

In this way, the whole spectrum of Rydberg states from $n = 2$ to the maximum visible P line is taken into account, even if the corresponding lines are not in the spectral region of measurement. This is essential because due to the spectrally far-reaching wings of the asymmetric Lorentzian line-shape, all P states influence the whole spectrum.

III. Results

The results of the fitting procedure can be grouped into three categories: 1. the parameters of the P lines (resonance energy, oscillator strength, linewidth, and asymmetry), 2. the parameters of the continuum absorption (band-edge shift, absorption strength, damping, and Urbach tail), and 3. parameters describing the fixed quantities (sample thickness, roughness, indirect absorption processes). Since the experiments have been performed with the same sample used in [49], we can take all parameters independent of the yellow P states (excitons and continuum) from Ref. 46.

We first discuss the pump-power dependence of the parameters for the P lines as obtained from fitting the transmission spectra of Fig. 1. This gives the oscillator strengths $\tilde{f}_n$ (Fig. 2a-b), the linewidths $\Gamma_n$ (Fig. 2c-d), the energies $E_n$ (Fig. 3), and the asymmetries $A_n$ (see Appendix I).

In our fitting strategy, we have to take into account several constraints. From many-body theory we expect no significant deviations of the oscillator strength from the law given by exciton theory: $f_n \propto (n^2 -1)/n^5$ [38]. However, due to the presence of charged impurities in the sample, we have also to consider a possible appearance of an Urbach tail accompanied by a reduction of oscillator strength. Since at high plasma densities, i.e., at high pump powers, the charges are efficiently screened and therefore of weak influence on the spectra, we start the fitting procedure at the highest pump power and continue to lower pump powers. Indeed, for pump powers above $100\,\mu W$ (Fig. 1, high power range (HPR)) the spectra can be fitted almost quantitatively by taking the oscillator strength fixed at the value expected from exciton theory (full colored lines in Fig. 2b). As is obvious from the spectra in Fig. 1 such a fit is not possible anymore for pump powers below 100 µW (low power range (LPR)). Near the band edge, there is a reduction in absorption (see black line in the data for $50\,\mu W$ as an example) that can be attributed to an Urbach-like tail. It turns out that a good fit can be obtained only by a reduction of the oscillator strength below that of exciton theory (see Fig. 2a). From absorption measurements without an additional pump, this phenomenon is already known (see e.g. Refs. 27 and 46) and has been explained as an effect of charged impurities [39]. It is reproduced here for pump powers below $2\,\mu\text{W}$, as can be seen in the behavior of the oscillator strengths (Fig. 2a), where the data sets are almost identical and show the characteristic additional decrease of oscillator strength above n=9. For pump powers between 5 and 50 $\mu\text{W}$ the dependence of oscillator strength on principal quantum number seems to be not systematic showing neither the variation at low nor that at high pump powers. To understand this, one has to treat the combined effects of impurities and EHP on the same footing. This requires further studies which are beyond the scope of this paper.

Surprisingly, the line broadenings $\Gamma_n$ (Fig. 2c-d) also show a different behavior in the different pump power ranges. In the HPR (Fig. 2d) the linewidths first decrease with increasing quantum number, reflecting the reduced coupling to phonons, but then start to increase again with a rate that depends on the pump power. The total linewidth may be split into two parts

$$\Gamma_{tot}(n) = \Gamma_{phon}(n) + \Gamma_{add}(n) \ , \tag{7}$$

where the phonon contribution $\Gamma_{phon}(n) \propto (n^2 - 1)/n^5$ is given in Figs. 2c and d as dashed black lines. The additional contribution shows a distinct dependence on principal quantum number $\Gamma_{add}(n) = C_{add}(n-1)^\zeta$ with $\zeta = 3.3 \pm 0.2$, see the full lines through the points. The strength factor $C_{add}$ obtained from a fit (full lines in Fig. 2d) obviously increases with increasing pump power (see inset in Fig. 2d). Therefore, one can speculate that the scattering process responsible for this broadening is related to the EHP. In Ref. 53 it was shown that scattering between different angular momentum states (here designated as usual with $l = 0$ (S), $l = 1$ (P), and $l = 2$ (D)) with the same principal quantum number by coupled plasmon-phonon modes results in a line broadening that scales with the quantum number as $(n-1)^{10/3}$ with a plasma-temperature dependent strength factor $C_{add} = 1.6 \rho_{eh} / \sqrt{T_{eh}/K}$ µeVµm$^3$ [54]. Accordingly, the strength factors directly allow deducing the ratios between the electron-hole densities and the square root of plasma temperature $\rho_{eh}/\sqrt{T_{eh}}$.

In contrast, in the LPR the linewidths show a monotonous decrease, which seems to saturate at high quantum numbers. The data can be reproduced quite well by adding a constant contribution $\Gamma_{add}(n) = \Gamma_r$, which amounts to about $6\,\mu\text{eV}$. As shown in [39] the features in the LPR (reduction of oscillator strength and existence of an Urbach tail) are a unique signature of the effects of charged impurities allowing us to conclude that in this range these spectra are predominantly impurity dominated. Therefore, an analysis of the linewidths and a comparison to theoretical results has been undertaken only for powers larger than $100\,\mu\text{W}$ (HPR), where we expect the influence of the EHP to be the dominant contribution.

Finally, the shift of the resonance energies $E_n$ with reference to a hypothetical "exact" energy of each P state in the form of a quantum defect formula is plotted in Fig. 3 for selected pump powers. Here we use a quantum defect formula in the two-parameter form
$E_n = -Ry/\left(n - \delta_1 - \delta_2/(n-\delta_1)^2 - \ldots\right)$ [49] with

$Ry = (87.3 \pm 0.2)$ meV, $\delta_1 = 0.225 \pm 0.02$, and $\delta_2 = -0.702 \pm 0.05$ allowing an accuracy of 0.6 µeV. While at low power ($< 50$ µW) no systematic deviation from the low-power results has been found, the measured red shift at higher pump powers follows the theoretically expected dependence on the quantum number given by $n^4$ [43] very well (see Eq. (9)), as shown by the solid lines.

The other quantity of interest obtained from the fit is the shift $\Delta$ of the apparent band edge $E_g$ which is indicated by the red arrows in Fig. 1 and as black squares in Fig. 4a. Here, it should be noted that the density of the charged impurities could be derived from the shift of the band edge at zero pump power using the results from Ref. 38 to be about $10^9$ cm$^{-3}$.

The asymmetries $A_n$ and the parameters of the Urbach-like tail are given for completeness in Appendix I (Figs. A2 and A3). As the latter originates from the effect of the residual charged impurities it is relevant only in the LPR, and therefore is of minor interest in this study.

## IV. Comparison with many-body plasma theory

We now compare the experimental results with those of the many-body plasma theory in the HPR. The parameters that determine the theoretical results are the electron-hole pair density $\rho_{eh}$ and the effective temperature $T_{eff}$ of the plasma. We assume that the carriers are in quasi-thermal equilibrium, which is justified by luminescence measurements [56] which show under various conditions (excitation energy, pump power) an exponential tail to higher photon energies in the luminescence from the EHP.

As the first input, we use the band-edge shift. Using the theory developed in Ref. 42, we can estimate the band-edge shift from

$$\Delta_{ehp}(\rho_{eh}, T_{eff}) = -6.92 \cdot 10^2 \mu eV \cdot \left(\rho_{eh} / \mu m^{-3}\right)^{1/4} \cdot (T_{eff} / K)^{1/4} \ . \tag{8}$$

This shows that one can determine only the product $\rho_{eh} \cdot T_{eh}$ from the experimental data. In lack of a rigorous theory for the effect of charged impurities in the presence of an EHP on the band-edge shift, we simply used the square root of the difference of the squares of the measured band-edge shift and the shift at zero pump power as input in these calculations. Although this might lead to some error at low powers, the high power data are not influenced.

The second input, the shift of the energy levels,

$$\Delta E_{th}(n, \rho_{eh}, T_{eh}) = -c_E n^4 \rho_{eh} (1 + b/T_{eh}^{1/4}) \ , \tag{9}$$

is also determined by $\rho_{eh}$ and $T_{eff}$ with $c_E = 1.5373 \cdot 10^{-2}$ µeVµm$^3$, and $b = 0.32$. To derive this equation we have to extend the calculation in [43] to a wider temperature range (see Appendix III). So we can determine from the energy shifts only the product $\rho_{eh}(1 + b/T_{eh}^{1/4})$. However, both relations together allow determining the density and the temperature of the EHP from the experiments by resolving the nonlinear temperature connection following from these relations. The values obtained are plotted as blue triangles in Fig. 4, where (b) shows the temperature and (c) the density. The densities obtained range from $\rho_{eh} \approx 10^{-3} \mu m^{-3}$ almost to $1 \mu m^{-3}$ and the temperatures vary between 5 and 20 K by changing the pump power from $20 \mu W$ to 5 mW.

As the third input we take the broadening of the linewidth, Eq. (7). Using again the band-edge shift, we can obtain a second independent set of densities and plasma temperatures, which are shown as the red dots in Figs. 4b and c. The agreement between both results is excellent. This is further substantiated by the agreement with the results obtained from the relaxation and cooling model (full lines) presented in Appendix II. A closer look at Fig. 4c reveals that the transition between LPR and HPR occurs at a pump power where the plasma density is one order of magnitude larger than that of the charged impurities ($10^9$ cm$^{-3}$), and one expects an efficient screening of these charges.

V. Failure of the Debye model

In order to account for the influence of many-body effects on the exciton states, the apparently most simple possibility is to replace the bare Coulomb potential by a statically screened one, e.g., the Debye potential (Eq. (1)), with the screening length given by Eq. (2). Then the shift of the band edge is given by Eq. (3), but with doubled value, as there are two species in the plasma. Simply solving the Schrödinger equation with this screened Coulomb potential, however, has long been known to yield an incorrect behavior of the bound-state energies in such a way that they are strongly blue-shifted towards the (unperturbed) band edge [15]. The subtraction of the single-particle shift as in [9] is only heuristically motivated [14]. A consistent description requires to include the many-body effects in the calculation of energy shifts on the same level of approximation in the single-particle and two-particle energies [43,45,46]. These calculations yield for the energy shifts

$$\Delta E_{thD}(n,l,\rho_{eh},T_{eff}) = -\alpha_{nl} \frac{2e_0^2}{\varepsilon_0 \varepsilon_{st}} \frac{\rho_{eh}}{k_B T_{eff}} \quad (10)$$

with the $\alpha_{nl}$ depending on the quantum numbers $n$ and $l$. From the calculations in [43], we extract for the $l=1$ (P) exciton states in Cu$_2$O $\alpha_{n1} = 26.5 \mu eV \mu m^3 / K \cdot n^{3/2}$. Note that the dependence on quantum number $n$ differs from the result of Refs. 9 and 57.

Inserting the parameters for the yellow excitons in Cu$_2$O (reduced electron-hole mass $\mu = 0.363 m_e$, $\varepsilon_r = 7.51$ [49]) into Eq. (3), we obtain for the band-edge shift

$$\Delta_{ehp}(\rho_{eh},T_{eff}) = c_D \cdot (\rho_{eh} / T_{eff})^{1/2} \quad (11)$$

with $c_D = -1.434 \cdot 10^3 \mu eV \cdot K^{1/2} \mu m^{-3/2}$.

Here, we see that both the shift of the band edge and of the energy levels depend only on the ratio $\rho_{\text{eh}}/T_{\text{eff}}$ albeit on different powers. This will provide a strong consistency check of the static screening model.

Figure 5a shows a comparison of the energy shifts of the P states as a function of the principal quantum number for different pump powers to the predictions of the Debye model. While results from the dynamically screened model certainly are a better match to the experimental results, the experimental accuracy does not allow us to unambiguously rule out the Debye model based solely on the energy shifts. However, the ratios of plasma density and temperature obtained from the slopes within the Debye model as blue dots in Fig. 5 differ by more than an order of magnitude from those determined from the experimental band-edge shifts (Fig. 4a) using Eq. (11), see red triangles in Fig. 5b. Note that both values, which for a consistent physical theory should be identical, differ by more than an order of magnitude. This clearly shows that the Debye model, if applied to the low-density EHP, gives an inadequate description. This fact casts considerable doubts on the application of the Debye model for other plasmas at low temperatures and densities as well.

## VI. Conclusions

To sum up, we have found two quite distinct regimes in the response of Rydberg excitons to an additionally present electron-hole plasma. For pump powers below $100\,\mu\text{W}$ (low power region), the effects are dominated by the influence of charged impurities leading to a reduction of the oscillator strength approaching the Mott transition, as observed earlier [33,37,39]. For pump powers above $100\,\mu\text{W}$ (high power range), however, we found the behavior predicted by many-body theory: (i) a clearly observable energy shift of the P absorption lines in the $\mu\text{eV}$ range, which agrees quantitatively with the predictions of many-particle theory on quantum number, plasma density, and temperature; (ii) a dependence of oscillator strength that follows the predictions of many-particle theory thus keeping the $(n^2-1)/n^5$ dependence of exciton theory [38]. In addition, we found an increase of the optical linewidth with the plasma density showing a characteristic dependence on quantum number. Such an effect does not occur in many-particle theory, which only gives a small additional contribution to the linewidth [44]. We have shown that this line broadening originates from the interaction of Rydberg excitons with coupled plasmon-phonon modes of the polar crystal $Cu_2O$, as recently predicted [53]. The plasma-induced line broadening then gives rise to an increasing overlap of the P lines with increasing quantum number thus leading to the background on which the lines sit. In a previous analysis [37] this has been erroneously interpreted as an Urbach-like tail reaching down in energy from the band edge, which in reality is not present in the high power range however. The many-body analysis of the energy shifts allows obtaining the plasma density and temperature independently, whereby the measured shift of the band edge is in quantitative agreement with the calculated shift.

Finally, we could show from our analysis that the standard Debye description for non-degenerate plasmas, which is generally thought to be a good approximation, is inadequate for the understanding of the presented experimental data.

**Acknowledgments**

The authors thank Peter Grünwald and Sjard Ole Krüger, Rostock, for helpful discussions. W.-D. K. thanks Ronald Redmer, Rostock, for hospitality. D.S. thanks the Deutsche Forschungsgemeinschaft for financial support (project number SE 2885/1-1), the Dortmund side acknowledges the support by the Deutsche Forschungsgemeinschaft through the International Collaborative Research Centre TRR160 (Project A8), and AS 459/3-1.

**References**

[1] I. Langmuir, *Oscillations in Ionized Gases*, P. Natl. Acad. Sci. USA **14**, 627 (1928).

[2] A. Piel, *Plasma physics: an introduction to laboratory, space, and fusion plasmas* (Springer, Berlin, Heidelberg, 2017).

[3] R. Hippler, H. Kersten, M. Schmidt, and K. H. Schoenbach (eds.), *Low Temperature Plasmas Vol I and II* (Wiley-VCH, Weinheim, 2008).

[4] R. Redmer, *Thermodynamic and transport properties of dense, low-temperature plasmas*, Phys. Rep. **282**, 35 (1997).

[5] H. Margenau and M. Lewis, *Structure of Spectral Lines from Plasmas*, Rev. Mod. Phys. **31**, 569 (1959).

[6] W. D. Kraeft, D. Kremp, W. Ebeling, and G. Röpke, *Quantum Statistics of Charged Particle Systems* (Akademie–Verlag, Berlin, 1986).

[7] D. Kremp, M. Schlanges, and W.D. Kraeft, *Quantum Statistics of Nonideal Plasmas*, Springer Series on Atomics, Optical, and Plasma Physics, Vol. 25 (Springer–Verlag, Berlin, 2005).

[8] P. Debye and E. Hückel, *Zur Theorie der Elektrolyte. I. Gefrierpunktserniedrigung und verwandte Erscheinungen*, Physik. Zeitschrift **24**, 185 (1923).

[9] G. Ecker and W. Weizel, *Zustandssumme und effektive Ionisierungsspannung eines Atoms im Inneren des Plasmas*, Annalen der Physik **452**, 126-140 (1956).

[10] W. D. Kraeft, K. Kilimann, and D. Kremp, *Theory of Nondegenerate two Component Systems with COULOMB Interaction*, Ann. Phys. (Leipzig) **29**, 177 (1973).

[11] W. D. Kraeft, K. Kilimann, and D. Kremp, *Quantum statistics of an electron-hole plasma*, Phys. Status Solidi B **72**, 461 (1975).

[12] K. Kilimann, W. D. Kraeft, and D. Kremp, *Lifetime and level shift of bound states in plasmas*, Phys. Lett. A **61**, 393 (1977).

[13] G. Röpke, K. Kilimann, D. Kremp, and W. D. Kraeft. *Two-particle energy shifts in low density non-ideal plasmas*, Phys. Lett. A **68**, 329 (1978).

[14] R. Zimmermann, K. Kilimann, W. D. Kraeft, D. Kremp, and G. Röpke, *Dynamical screening and self-energy of excitons in the electron-hole plasma*, Phys. Status Solidi B **90**, 175 (1978).

[15] W. Ebeling, W. D. Kraeft, and D. Kremp, *Theory of Bound States and Ionization Equilibrium in Plasmas and Solids* (Akademie–Verlag, Berlin, 1976).

[16] O. Ciricosta, *et al.*, *Detailed model for hot-dense aluminum plasmas generated by an X-ray free electron laser*, Phys. Plasmas **23**, 022707 (2016).

[17] S. Vinko, *et al.*, *Creation and diagnosis of a solid-density plasma with an X-ray free-electron laser*, Nature **482**, 59 (2012).

[18] B. J. B. Crowley and G. Gregori, *Quantum theory of Thomson scattering*, High Energ. Dens. Phys. **13**, 55-83 (2014).

[19] C. Lin, G. Röpke, W.-D. Kraeft, and H. Reinholz, *Ionization-potential depression and dynamical structure factor in dense plasmas*, Phys. Rev. E **96**, 013202 (2017).

[20] G. Röpke, D. Blaschke, T. Döppner, C. Lin, W.-D. Kraeft, R. Redmer, and H. Reinholz, *Ionization potential depression and Pauli blocking in degenerate plasmas at extreme densities*, Phys. Rev. E **99**, 033201 (2019).

[21] H. Ehrich and H. J. Kusch, *Die Erniedrigung der Ionisationsenergie des CII-Ions in einem Hochdruckplasma*, Z. Naturforsch. A **29**, 810 (1974).

[22] D. J. Hoarty, P. Allan, S. F. James, C.R. D. Brown, L. M. R. Hobbs, M. P. Hill, J. W. O. Harris, J. Morton, M. G. Brookes, R. Shepherd, J. Dunn, H. Chen, E. von Marley, P. Beiersdorfer, H. K. Chung, R. W. Lee, G. Brown, J. Emig, *Observations of the Effect of Ionization-Potential Depression in Hot Dense Plasma*, Phys. Rev. Lett. **110**, 265003 (2013).

[23] P. Braun-Munzinger and J. Stachel, *The quest for the quark-gluon plasma*, Nature **448,** 302–309 (2007).

[24] G. Röpke, D. Blaschke, and H. Schulz, *Dissociation kinetics and momentum-dependent J/Ψ suppression in a quark-gluon plasma*, Phys. Rev. D **38**, 3589 (1988).

[25] C. Klingshirn and H. Haug, *Optical properties of highly excited direct gap semiconductors*, Phys. Rep. **70**, 315 (1981).

[26] C. Klingshirn, *Semiconductor Optics*, 4th ed. (Springer, Berlin, Heidelberg, 2012).

[27] H. Haug and S. W. Koch, *Quantum Theory of Optical and Electronic Properties of Semiconductors* (World Scientific, Singapore 2004).

[28] H. Haug and S. Schmitt-Rink, *Electron theory of the optical properties of laser-excited semiconductors*, Prog. Quant. Electr. **9**, 3 (1984).

[29] R. Zimmermann, *Many-Particle Theory of Highly Excited Semiconductors*, Teubner–Texte zur Physik Bd. 18 (BSB B. G. Teubner Verlagsgesellschaft, Leipzig, 1988).

[30] D. Semkat, F. Richter, D. Kremp, G. Manzke, W.-D. Kraeft, and K. Henneberger, *Ionization equilibrium in an excited semiconductor: Mott transition versus Bose-Einstein condensation*, Phys. Rev. B **80**, 155201 (2009).

[31] D. Semkat, F. Richter, D. Kremp, G. Manzke, W.-D. Kraeft, and K. Henneberger, *Mott transition in electron-hole plasmas*, J. Phys.: Conf. Ser. **220**, 012005 (2010).

[32] G. Manzke, D. Semkat, and H. Stolz, *Mott transition of excitons in GaAs-GaAlAs quantum wells*, New J. Phys. **14**, 095002 (2012).

[33] T. Kazimierczuk, D. Fröhlich, S. Scheel, H. Stolz, and M. Bayer, *Giant Rydberg excitons in the copper oxide $Cu_2O$*, Nature **514**, 343 (2014).

[34] M. Aßmann and M. Bayer, *Semiconductor Rydberg Physics*, Adv. Quantum Techn. **3**, 1900134 (2020).

[35] J. Heckötter, D. Janas, R. Schwartz, M. Aßmann, and M. Bayer, *Experimental limitation in extending the exciton series in $Cu_2O$ towards higher principal quantum numbers*, Phys. Rev. B **101**, 235207 (2020).

[36] M. A. M. Versteegh, *et al.*, *Giant Rydberg excitons in $Cu_2O$ probed by photoluminescence excitation spectroscopy*, arXiv:2105.07942v1 (2021).

[37] J. Heckötter, M. Freitag, D. Fröhlich, M. Aßmann, M. Bayer, P. Grünwald, F. Schöne, D. Semkat, H. Stolz, and S. Scheel, *Rydberg Excitons in the Presence of an Ultralow-Density Electron-Hole Plasma*, Phys. Rev. Lett. **121**, 097401 (2018).

[38] R. J. Elliott, *Intensity of Optical Absorption by Excitons*, Phys. Rev. **108,** 1384 (1957).

[39] S. O. Krüger, H. Stolz, and S. Scheel, *Interaction of charged impurities and Rydberg excitons in cuprous oxide*, Phys. Rev. B **101**, 235204 (2020).

[40] V. Walther and Th. Pohl, *Plasma-Enhanced Interaction and Optical Nonlinearities of $Cu_2O$ Rydberg Excitons*, Phys. Rev. Lett. **125**, 097401 (2020).

[41] R. Schwartz, N. Naka, F. Kieseling, and H. Stolz, *Dynamics of excitons in a potential trap at ultra-low temperatures: paraexcitons in $Cu_2O$*, New J. Phys. **14,** 023054 (2012).

[42] D. Semkat, S. Sobkowiak, F. Schöne, H. Stolz, Th. Koch, and H. Fehske, *Multicomponent exciton gas in cuprous oxide: cooling behaviour and the role of Auger decay*, J. Phys. B: At. Mol. Opt. Phys. **50**, 204001 (2017).

[43] D. Semkat, H. Fehske, and H. Stolz, *Influence of electron-hole plasma on Rydberg excitons in cuprous oxide*, Phys. Rev. B **100**, 155204 (2019).

[44] D. Semkat, H. Fehske, and H. Stolz, *Quantum many-body effects on Rydberg excitons in cuprous oxide*, Eur. Phys. J. Spec. Top. **230**, 947 (2021).

[45] J. Seidel, S. Arndt, and W. D. Kraeft, *Energy spectrum of hydrogen atoms in dense plasmas*, Phys. Rev. E **52**, 5387 (1995).

[46] S. Arndt, W. D. Kraeft, and J. Seidel, *Two-particle energy spectrum in dense electron-hole plasmas*, Phys. Stat. Sol. B **194**, 601 (1996).

[47] J. Heckötter, M. Freitag, D. Fröhlich, M. Aßmann, M. Bayer, M. A. Semina, and M. M. Glazov, *High-resolution study of the yellow excitons in $Cu_2O$ subject to an electric field*, Phys. Rev. B **95**, 035210 (2017).

[48] Actually, it turns out that the maximum observable $n$ is slightly dependent on the position of the laser spot on the sample due to the sensitivity on impurity concentration that varies over the sample, but also on surface quality (see Ref. 36) and temperature (see Ref. 35).

[49] H. Stolz, R. Schwartz, J. Heckötter, M. Aßmann, D. Semkat, S. O. Krüger, and M. Bayer, *Coherent transfer matrix analysis of the transmission spectra of Rydberg excitons in cuprous oxide*, Phys. Rev. B **104**, 035206 (2021).

[50] F. Schöne, H. Stolz, and N. Naka, *Phonon-assisted absorption of excitons in $Cu_2O$*. Phys. Rev. B, **96**, 115207 (2017).

[51] T. Ito, T. Kawashima, H. Yamaguchi, T. Masumi, and S. Adachi, *Optical properties of $Cu_2O$ studied by spectroscopic ellipsometry*, J. Phys. Soc. Jpn. **67**, 2125 (1998).

[52] Obtained from extrapolating the exciton energies for a principal quantum number going to infinity (for details see [49,55]).

[53] H. Stolz and D. Semkat, *Scattering of Rydberg excitons by phonon-plasmon modes*, J. Phys.: Condens. Matter **33**, 425701 (2021).

[54] At variance with the results of Ref. 53, we need here the half-width at half maximum of the Lorentz line and not $\hbar/T_1$ with $T_1$ the scattering time. Considering that in $Cu_2O$ there are two LO phonons, we should also double the matrix elements.

[55] F. Schöne, S. O. Krüger, P. Grünwald, H. Stolz, S. Scheel, M. Aßmann, J. Heckötter, J. Thewes, D. Fröhlich, and M. Bayer, *Deviations of the exciton level spectrum in $Cu_2O$ from the hydrogen series*, Phys. Rev. B **93**, 075203 (2016).

[56] M. Takahata and N. Naka, *Photoluminescence properties of the entire excitonic series in $Cu_2O$*, Phys. Rev. B **98**, 195205 (2018).

[57] F. A. Rogers, H. C. Graboske Jr., and D. J. Harwood, *Bound Eigenstates of the Static Screened Coulomb Potential*, Phys. Rev. A **1**, 1577 (1976).

[58] Y. Toyozawa, *Interband effect of lattice vibrations in the exciton absorption spectra*, J. Phys. Chem. Solids **25**, 59 (1964).

[59] F. Schweiner, J. Main, and G. Wunner, *Linewidths in excitonic absorption spectra of cuprous oxide*, Phys. Rev. B **93**, 085203 (2016).

[60] J. Thewes, J. Heckötter, T. Kazimierczuk, M. Aßmann, D. Fröhlich, M. Bayer, M. A. Semina, and M. M. Glazov, *Observation of High Angular Momentum Excitons in Cuprous Oxide*, Phys. Rev. Lett. **115**, 027402 (2015).

[61] D. P. Trauernicht and J. P. Wolfe, *Drift and diffusion of paraexcitons in $Cu_2O$: Deformation-potential scattering in the low-temperature regime*, Phys. Rev. B **33**, 8506 (1986).

[62] B. K. Ridley, *Quantum Processes in Semiconductors*, 5th edition (Oxford University Press, Oxford, 2013).

[63] Y. Toyozawa, *Polarons in Ionic Crystals and Polar Semiconductors,* ed. J. Devreese (North Holland, Amsterdam, 1972), p. 1.

[64] N. Naka, I. Akimoto, M. Shirai, and K. I. Kan'no, *Time-resolved cyclotron resonance in cuprous oxide*, Phys. Rev. B **85**, 035209 (2012).

[65] M. French, R. Schwartz, H. Stolz, and R. Redmer, *Electronic band structure of Cu2O by spin density functional theory*, J. Phys.: Condens. Matter **21**, 015502 (2009).

[66] G. M. Kavoulakis and G. Baym, *Auger decay of degenerate and Bose-condensed excitons in $Cu_2O$*, Phys. Rev. B **54**, 16625 (1996).

[67] A. Farenbruch, D. Fröhlich, H. Stolz, D. R. Yakovlev, and M. Bayer, *Second-harmonic generation of blue series excitons and magnetoexcitons in $Cu_2O$*, Phys. Rev. B **104**, 075203 (2021).

[68] A. Griffin, T. Nikuni, and E. Zaremba, *Bose-Condensed Gases at Finite Temperature,* (Cambridge University Press, Cambridge, 2009).

[69] S. Sobkowiak, D. Semkat, and H. Stolz, *Hydrodynamic description of trapped ultracold paraexcitons in $Cu_2O$*, Phys. Rev. B **91**, 075209 (2015).

Appendix I: Transmission spectra and fitting results

(a) Examples of a fit of the transmission spectra

In Fig. A1 we show a detailed comparison of the experimental transmission spectrum (red line) for a pump power of 0.2 µW (panel a) and 1000 µW (panel b) with a fit using the CTM method (for details of the method we refer to Ref. 46). In addition to the theoretical transmission spectrum (blue line), we also show the theoretical reflection spectrum (magenta line). This shows a considerable spectral variation around the P lines, which cannot be neglected. The difference between experimental and theoretical transmission spectra is given by the green line. The most prominent deviations are due to absorption into higher angular momentum states. The sharp structures above the P lines are the well-known $l = 3$ (F) states [60]. Other structures are due to $l = 2$ (D) states becoming allowed due to the electric field of ionized residual impurities [39].

(b) Results for the asymmetry parameters

It is well known that the absorption lines of the yellow P states are sitting on top of a phonon background. Because both processes are leading to the same final state, we observe for the P lines an asymmetric Lorentzian line shape given by Eq. (4) [58]. One should note that at present a quantitative calculation of the dependence of the asymmetry on principal quantum number is not available [59]. We therefore only show the results for the asymmetry parameters obtained from the fit of the transmission spectra in Fig. A2.

(c) Parameters of the Urbach tail

To ensure a smooth transition from the continuum to the Urbach tail, the sharp spectral dependence given by Eq. (1.3) has to be convoluted by a Lorentzian with the same width as that for the continuum. The result is

$$\varepsilon_{UT}(\hbar\omega, E_{ur}) = -a_c \frac{i}{2\pi} \exp\left(\frac{-i\gamma_c + z}{E_{ur}}\right) \left[ \text{Ei}\left(\frac{-i\gamma_c + z}{E_{ur}}\right) - \exp\left(\frac{2i\gamma_c}{E_{ur}}\right) \text{Ei}\left(\frac{i\gamma_c + z}{E_{ur}}\right) \right] \quad (A1)$$

with $z = -E_g + \hbar\omega$.

In the actual fit, we assumed a two-component expression for the Urbach tail

$$\varepsilon_{UTtot}(\hbar\omega) = \left( A_{ur}\varepsilon_{UT}(\hbar\omega, E_{ur1}) + (1 - A_{ur})\varepsilon_{UT}(\hbar\omega, E_{ur2}) \right) \Phi(E_g - \hbar\omega) \quad (A2)$$

With $A_{ur}$ being a weighting factor normalized to unity.

The parameters of the Urbach tail obtained from the fitting procedure are shown in Fig. A3. Note that only for pump powers below 100 µW the Urbach tail is relevant in the spectra. So we do not discuss it further here.

Appendix II: Relaxation and cooling model

(a) Rate model

Here we discuss the rate model by which the relaxation processes of the different species are described.

In the past years, we have developed a rate model for analyzing continuous-wave and pulsed time-resolved measurements of 1S yellow excitons in $Cu_2O$ [34,35], which can be applied also for the excitation scenario in this paper, as one primarily excites only yellow 1S orthoexcitons at large wave vectors. These relax by phonon emission towards the band minimum and convert into 1S paraexcitons. Both decay by a bimolecular Auger process into electron-hole pairs. This limits the relaxation and conversion processes to:

  i. Acoustical and optical phonon scattering of 1S excitons.
  ii. Relaxation of eh pairs into excitons. We consider this as a bimolecular process, mediated either by $\Gamma_3^-$ phonons or as a direct capture into Rydberg exciton states with high principal quantum numbers.
  iii. Trapping of all species at defects and due to the small thickness of the sample also surface recombination.
  iv. Conversion of ortho- into paraexcitons.
  v. Auger decay of ortho- and paraexcitons.

Of course, other relaxation processes are imaginable, like Auger recombination of electrons or holes. However, signatures for such processes, which would scale with the third power of the carrier concentration, have not been found.

The resulting system of rate equations is then given by

$$\frac{dn_{\text{Oex}}}{dt} = \eta_L G(t) - \Gamma_{\text{relO1S}} n_{\text{Oex}} - \Gamma_{\text{O1Snr}} n_{\text{Oex}} - \Gamma_{\text{OP}} n_{\text{Oex}} - 2 a_{\text{OO}} n_{\text{Oex}}^2 - a_{\text{OO}} n_{\text{Oex}} n_{\text{O1S}} - a_{\text{OP}} n_{\text{Oex}} n_{\text{P1S}} + \Gamma_{raug} n_{\text{aug}}^2, \quad \text{(A3)}$$

$$\frac{dn_{\text{O1S}}}{dt} = \Gamma_{\text{relO1S}} n_{\text{Oex}} - \Gamma_{\text{O1Snr}} n_{\text{O1S}} - \Gamma_{\text{OP}} n_{\text{O1S}} - 2 a_{\text{OO}} n_{\text{O1S}}^2 - a_{\text{OP}} n_{\text{O1S}} n_{\text{P1S}} - a_{\text{OO}} n_{\text{Oex}} n_{\text{O1S}}, \quad \text{(A4)}$$

$$\frac{dn_{\text{P1S}}}{dt} = \Gamma_{\text{OP}} (n_{\text{O1S}} + n_{\text{Oex}}) \beta_D - \Gamma_{\text{P1Snr}} n_{\text{P1S}} - 2 a_{\text{PP}} n_{\text{P1S}}^2 - (a_{\text{OP}} n_{\text{Oex}} n_{\text{P1S}} + a_{\text{OP}} n_{\text{O1S}} n_{\text{P1S}}) \beta_D, \quad \text{(A5)}$$

$$\frac{dn_{\text{aug}}}{dt} = a_{\text{PP}} n_{\text{P1S}}^2 \beta_D + a_{\text{OO}} n_{\text{O1S}}^2 + a_{\text{OO}} n_{\text{Oex}} n_{\text{O1S}} + a_{\text{OP}} n_{\text{Oex}} n_{\text{P1S}} \beta_D + a_{\text{OP}} n_{\text{O1S}} n_{\text{P1S}} \beta_D$$
$$- \Gamma_{raug} n_{\text{aug}}^2 - \Gamma_{augnr} n_{\text{aug}}. \quad \text{(A6)}$$

The rate constants are given in Table A1.

| Rate constant | Meaning | Value | Ref. |
|---|---|---|---|
| $\Gamma_{rOO1S}$ | Relaxation rate into 1S orthoexciton ground state | 100/ns | 34 |
| $\Gamma_{OP}$ | Ortho-para conversion rate | 0.2/ns | 34 |
| $a_{OO}$ | Ortho-ortho Auger rate | $0.67 \cdot 10^{-4}$ μm$^3$/ns | 34 |
| $a_{OP}$ | Ortho-para Auger rate | $a_{OO}/4$ | 35 |
| $a_{PP}$ | Para-para Auger rate | $4 \cdot 10^{-6}$ μm$^3$/ns | 34 |
| $\Gamma_{O1Snr}$ (*) $\Gamma_{P1Snr}$ | Non-radiative decay rate of ortho- and paraexcitons | 1/500ns | fit |
| Fraction of absorbed photons $\alpha_O$ | Absorption of orthoexcitons due to phonon process | 0.2 | [49] |
| $\Gamma_{augnr}$ (*) | Non-radiative decay rate of hot Auger generated electrons and holes | 1/500ns | Fit |
| $\Gamma_{raug}$ (*) | Recombination rate of hot Auger generated eh pairs into excitons | 1 μm$^3$/ns | Fit |

Table A1: Parameters of the rate model. The parameters marked by (*) are either sample dependent or not known, so have to be obtained from a fitting procedure.

To calculate the excitation densities from the pump laser power, we consider an excitation spot with a Gaussian intensity

$$I(\rho) = I_0 \exp\left(-\left(\frac{\rho}{\rho_0}\right)^2\right) \quad (A7)$$

with waist parameter $\rho_0 = 180$μm, which is related to the FWHM beam diameter $2R_0$ by $R_0 = \sqrt{\ln 2}\,\rho_0$. The beam power is given by $P_L = \pi \rho_0^2 I_0$. Since the probe beam has only a diameter of $\rho_P = 90$μm, it senses only the central part. The overlap parameter defined as

$$c_{ovl}(\rho_0, \rho_P) = \frac{2}{\rho_P^2} \int_0^\infty \rho \exp\left(-\left(\frac{\rho}{\rho_0}\right)^2\right) \exp\left(-\left(\frac{\rho}{\rho_P}\right)^2\right) d\rho \quad (A8)$$

is 0.923, which allows for linear processes to assume a rectangular beam profile with diameter $\rho_0$ and power $c_{ovl}P_L$. We note that the absorbed fraction of laser power $\eta_L$ cannot be obtained in the usual way from Lambert's law, but must follow from an analysis of the transmission by the CTM method as shown in [49]. This gives for the sample at the excitation photon energy $\eta_L = 0.2$, which includes all loss by scattering and reflection of the sample and also of the quartz windows.

To include diffusion, which however affects only the 1S paraexcitons due to their long lifetime and high diffusivity ($D > 1000$ cm$^2$/s [61]), we introduce a factor $\beta_D = R_0^2/(R_0^2 + D/\Gamma_{P1Snr})$ in all paraexciton out-scattering rates. $G(t)$ is given by

$$G(t) = c_{ovl}P_L \frac{1}{\pi\rho_0^2 \cdot E_L} \text{ns} \cdot f(t) \ . \tag{A9}$$

For a cw pump we set $f(t) = 1$. $E_L$ is the laser photon energy.

In Fig. A4 a typical result for the dependence of the species' concentrations on pump power is shown. Obviously, the rate model allows reproducing the experimentally found plasma density quite accurately.

### (b) Cooling of an electron-hole plasma

The properties of an EHP strongly depend on its density and temperature. As information on these quantities is not easily available directly, one has to model the relaxation behavior of the primarily excited electrons and holes applying kinetic equations. This approach has a long tradition in semiconductor physics (see, e.g., Ref. 56).

The first step in such a treatment is a dedicated description of the generation and scattering processes the carriers undergo. Here one has to consider the following processes:

1. scattering of carriers by phonons,
2. carrier-carrier scattering by Coulomb interaction,
3. scattering and trapping of carriers by impurities,
4. recombination of carriers into excitons.

In addition, one has to know the distribution of the carriers over the band states excited by optical absorption or by the Auger-like decay process of excitons.

#### 1. Scattering by phonons

Here we have to distinguish scattering by acoustical phonons (deformation potential ac) and optical phonons (Fröhlich interaction, LO).

For the ac-process, the scattering rates are given in Ref. 56. In the limit of low carrier concentration and low temperatures we have

$$\Gamma_{x,ph}(k) = \frac{D^2}{2\rho u_L} \frac{1}{(2\pi)} \frac{m^* \kappa^2}{\hbar^2 k} \frac{8}{3}\left(\sqrt{\frac{E}{E_{ph}}} - 1\right)^3 = \frac{2}{3\pi} \frac{D^2}{\rho v_S} \frac{m^* \kappa^2}{\hbar^2} \frac{1}{\sqrt{\frac{E}{E_{ph}}}}\left(\sqrt{\frac{E}{E_{ph}}} - 1\right)^3, \quad (A10)$$

with deformation potential D, sound velocity $u_L$, $E_{ph} = \frac{\hbar^2}{2m^*}\kappa^2$, and $\kappa = 2m^* u_L/\hbar$.

For LO scattering we have [59]

$$\Gamma_{c,LOi} = \frac{e_0^2}{\varepsilon_0}\left(\frac{1}{\varepsilon_i^*}\right)\frac{1}{4\pi\hbar^2}\sqrt{\frac{m^*\hbar\Omega_{LOi}}{2}}\frac{1}{\sqrt{E(k)/\hbar\Omega_{LOi}}}\Theta(E(k) - \hbar\Omega_{LOi})\ln\left[\frac{1+\sqrt{1-\hbar\Omega_{LOi}/E(k)}}{1-\sqrt{1-\hbar\Omega_{LOi}/E(k)}}\right]. \quad (A11)$$

Here $\hbar\Omega_{LOi}$ denote the different phonon energies, $\varepsilon_i^*$ denotes the effective dielectric constants for the different phonon modes given as

$$\frac{1}{\varepsilon_i^*} = \frac{1}{\varepsilon_i^{up}} - \frac{1}{\varepsilon_i^{low}}, \quad (A12)$$

where $\varepsilon_i^{up,low}$ denote the dielectric constants above and below the phonon mode which can be calculated from Toyozawa's rule [63]

$$\frac{\varepsilon(\omega)}{\varepsilon_\infty} = \prod_i \frac{(\omega_{LOi}^2 - \omega^2)}{(\omega_{TOi}^2 - \omega^2)}. \quad (A13)$$

For $Cu_2O$ the corresponding values are [41]

$$\begin{aligned} E_{TO1} &= 18.8 \text{ meV} & E_{LO1} &= 19.1 \text{ meV} \\ E_{TO2} &= 78.5 \text{ meV} & E_{LO2} &= 82.1 \text{ meV} \end{aligned}, \quad (A14)$$

from which we have to determine the effective $\varepsilon_i^*$. We use

$$\varepsilon_{stat} = 7.37, \quad \varepsilon_{int} = 7.14, \quad \varepsilon_\infty = 6.53 \quad (A15)$$

resulting in $\varepsilon_1^* = 233$, $\varepsilon_2^* = 76$. As electron and hole mass we use the polaron masses $m_c^* = 0.985 m_e$, $m_v^* = 0.575 m_e$ [64].

In $Cu_2O$ the uppermost valence band has a significant non-parabolicity [65]. To take this approximately into account, one may divide the valence band into two sections. The part at low wave vectors has the usual mass given above, while the upper part corresponds to a mass of $m_{v2}^* = 3.5 m_e$, the energy difference between the bands is $\Delta_7 = 70$ meV. If initially the holes are excited with high kinetic energies, they will follow the upper band and as soon as they reach the energies below the crossing point at 85 meV, they switch to the lower band.

In Fig. A5 the scattering rates of the electrons (part a) and of the holes (part b) are plotted. We show both the rates for the lower and for the upper band. Fortunately for the upper band scattering by LO2 phonons is dominant, so we need to consider only this process. Due to the low particle concentration, the carrier-carrier scattering is very small.

Accordingly, the carriers reach very quickly the energies where scattering with LO2 modes is no longer possible at low temperatures.

2. Excitation processes

In the experiments to be analyzed the carriers are excited only by Auger decay of yellow 1S ortho- and paraexcitons [41,42,66]. This can proceed via a direct process (type I), whereby mono-energetic electrons and holes are excited as determined by momentum and energy conservation. The carriers then relax down via Fröhlich scattering. However, in the literature, there is general agreement that the phonon-assisted Auger process (type II) is strongest (see Ref. 41 and references therein). This proceeds predominantly via the blue exciton [42,67] so that we have the following scenario: The electrons are excited in the $\Gamma_8^-$ band, from which they relax to the bottom. From there they are scattered by interband scattering via the odd-parity optical phonon (the dominant process is that with the $\Gamma_3^-$ mode [42]) to the $\Gamma_6^+$ conduction band with a kinetic energy of about 440 meV (difference of band edge minus phonon energy), from where the relaxation begins with LO2 scattering, the rate of which is very large (see Fig. 5a). In contrast, the holes are excited to all valence band states only restricted by energy and momentum conservation. The starting energies for slow cooling are therefore for electrons 30 meV and for holes 40 meV.

3. Hydrodynamic cooling model

For the following hydrodynamic model for carrier cooling, we assume that

a. the system is homogeneous,

b. the carriers are in quasi-equilibrium with effective temperature T and chemical potential $\mu$ so that the carrier distribution is given by a Fermi distribution

$$f(\vec{k}) = \frac{1}{\exp\left(-\left(\frac{1}{k_B T}\right)\left(\frac{\hbar^2 k^2}{2m_i^*} - \mu\right)\right) + 1} \quad . \tag{A16}$$

Under these assumptions, the quantum Boltzmann equation can be solved by considering only the zeroth moment (carrier density) and the second moment (energy density) [68,69].

The equation for the zeroth moment becomes the usual rate equation

$$\frac{dn}{dt} = \Gamma_{\text{X-ph}}^{(0)} + \Gamma_{\text{E,V}}^{(0)} \quad . \tag{A17}$$

The second moment gives the equation for the total energy density

$$\frac{dE}{dt} = \Gamma_{\text{X-ph}}^{(2)} + \Gamma_{\text{E-V}}^{(2)} \quad . \tag{A18}$$

The moments for scattering by phonons for a parabolic band are [69]

$$\Gamma_{\text{X-ph}}^{(n)} = -\frac{2\pi}{\hbar}V\int\frac{d^3k}{(2\pi)^3}\int\frac{d^3k'}{(2\pi)^3}\left|M_{ph}(\vec{k}'-\vec{k})\right|^2\Phi_n(\vec{k})\left\{\left[f_{\vec{k}}(1+f_{\vec{k}-\vec{k}'}^{ph})(1-f_{\vec{k}'})\right.\right.$$
$$\left.-(1-f_{\vec{k}})f_{\vec{k}-\vec{k}'}^{ph}f_{\vec{k}'}\right]\delta(\varepsilon_{\vec{k}}-\varepsilon_{\vec{k}'}-\hbar\omega_i^{ph})+\left[f_{\vec{k}}f_{\vec{k}'-\vec{k}}^{ph}(1-f_{\vec{k}'})\right.$$
$$\left.\left.-(1-f_{\vec{k}})(1+f_{\vec{k}'-\vec{k}}^{ph})f_{\vec{k}'}\right]\delta(\varepsilon_{\vec{k}}-\varepsilon_{\vec{k}'}+\hbar\omega_i^{ph})\right\}$$

(A19)

where $M_{ph}(\vec{k}'-\vec{k})$ is the carrier phonon scattering matrix element [62] and $\Phi_n(k)$ denotes $\Phi_0(k)=1$, $\Phi_1(k)=\hbar\vec{k}$, $\Phi_2(k)=\hbar^2k^2/2m$. The phonon distribution is given by (with the lattice temperature $T_{\text{phon}}$ assumed to be constant)

$$f_q^{ph} = \frac{1}{\exp\left(\frac{\hbar u_L q}{k_B T_{\text{phon}}}\right)-1} \quad . \tag{A20}$$

For scattering by acoustical phonons via deformation-potential one obtains [62]

$$\Gamma_{\text{X-ph}}^{(2)} = \frac{D^2 m^2}{2\rho(2\pi)^3}\frac{k_B T}{\hbar^3}\int dq\, q^3 (f_q^{ph}-f_{\Omega(q)}^B)\left(\ln\left[\frac{e^{-\beta\left(\frac{\hbar^2}{8m}(q-\kappa)^2-\mu\right)}+1}{e^{-\beta\left(\frac{\hbar^2}{8m}(q+\kappa)^2-\mu\right)}+1}\right]\right) . \tag{A21}$$

For scattering with optical phonons via the Fröhlich interaction we can analogously derive

$$\Gamma_{\text{X-ph}}^{(2)} = \frac{C_i\hbar\Omega_{\text{LOi}}}{(2\pi)^3}\frac{m}{2\hbar^3}\frac{2mk_B T}{\hbar^2}(f_{\Omega_{\text{LO}}}^{ph}-f_{\Omega_{\text{LO}}}^B)\int dq\, q^{-1}\left(\ln\left[\frac{e^{-\beta\left(\frac{\hbar^2}{8m}(q-\kappa_0^2/q)^2-\mu\right)}+1}{e^{-\beta\left(\frac{\hbar^2}{8m}(q+\kappa_0^2/q)^2-\mu\right)}+1}\right]\right), \tag{A22}$$

with the matrix element

$$M_{LOi}^2 = C_i/V|\vec{k}-\vec{k}'|^2, \quad C_i = \frac{e_0^2}{2\varepsilon_0}\hbar\Omega_{LOi}\left(\frac{1}{\varepsilon_i^*}\right) . \tag{A23}$$

The zeroth and second moment for excitation and decay processes are simply given as

$$\begin{aligned}\Gamma_{exc}^{(0)} &= G(t), & \Gamma_{dec}^{(0)} &= -\Gamma_{eh}n(t), & \Gamma_{rec}^{(0)} &= -\Gamma_{rc}n(t)^2\\ \Gamma_{exc}^{(2)} &= E_{exc}(t), & \Gamma_{dec}^{(2)} &= -\Gamma_{eh}E(t), & \Gamma_{rec}^{(2)} &= -\Gamma_{rc}n\cdot E(t)\end{aligned} \tag{A24}$$

with $E_{exc} = E_{\text{start}}\cdot G_0$, where $E_{\text{start}}$ is the initial excess energy as provided by e.g. the Auger process.

$E(t)$ is the total energy of the plasma, which is given by

$$E = \int\frac{d^3k}{(2\pi)^3}\frac{\hbar^2k^2}{2m}f_{\vec{k}} \quad . \tag{A25}$$

The stationary solution of the hydrodynamic equations is obtained from

$$\frac{dn}{dt} = 0 = G_0 - \Gamma_{eh} n_0 - \Gamma_{rc} n_0^2 \qquad . \tag{A26}$$

$$\frac{dE}{dt} = 0 = \Gamma^{(2)}_{X\text{-ph}}(n_0,T_0) + E_{exc} - \Gamma_{eh} E(n_0,T_0) - \Gamma_{rc} n_0 \cdot E(n_0,T_0) \tag{A27}$$

Equation (A26) can be solved directly giving

$$n_0 = -\frac{\Gamma_{eh}}{2\Gamma_{rc}} + \sqrt{\left(\frac{\Gamma_{eh}}{2\Gamma_{rc}}\right)^2 + \frac{G_0}{\Gamma_{rc}}} \quad . \tag{A28}$$

Equation (A27) gives

$$E(n_0,T_0) = \frac{\Gamma^{(2)}_{X\text{-ph}}(n_0,T_0) + E_{exc}}{\Gamma_{eh} + \Gamma_{rc} n_0} \quad , \tag{A29}$$

which is an implicit relation for $T_0$ that is solved numerically.

In these equations, the zeroth and second moments of the creation and destruction of Auger created eh pairs have to be inserted. Since it seems impracticable to incorporate the complete rate model into the hydrodynamic equations, we try to approximate the whole system by taking into account only the relevant particles, the 1S paraexcitons, and the EHP. These would follow the following simplified rate equations

$$\frac{dn_{P1S}}{dt} = \alpha_S G(t) - \Gamma_{0P} n_{P1S} - 2 a_{0PP} n_{P1S}^2 \quad , \tag{A30}$$

$$\frac{dn_{aug}}{dt} = a_{0PP} n_{P1S}^2 - \Gamma_{0augnr} n_{aug} - \Gamma_{0raug} n_{aug}^2 \quad . \tag{A31}$$

Here the rates are effective rates that have to be adjusted to reproduce the results from the extended rate model.

The stationary solutions are easily obtained as

$$\begin{aligned} n_{0P}(P_L) &= -\frac{\Gamma_{0P}}{4 a_{0PP}} + \sqrt{\left(\frac{\Gamma_{0P}}{4 a_{0PP}}\right)^2 + \frac{G_0(P_L)}{2 a_{0PP}} \alpha_{0eff}} \\ n_{0aug}(P_L) &= -\frac{\Gamma_{0augnr}}{2\Gamma_{0raug}} + \sqrt{\left(\frac{\Gamma_{0augnr}}{2\Gamma_{0raug}}\right)^2 + \frac{a_{0PP}}{2\Gamma_{0raug}} n_{0P}(P_L)^2} \end{aligned} \quad . \tag{A32}$$

As an example, in Fig. A4 the modeling for the scenario described in the main paper is shown. The following parameters give the best approximation

$$\Gamma_{0P} = 2.67 \cdot 10^{-3} / ns, \quad a_{0PP} = 3 \cdot 10^{-6} \mu m^3 / ns,$$
$$\alpha_{0eff} = 2.15 \cdot 10^{-9}, \quad \Gamma_{0augnr} = \frac{1}{150 ns}, \quad \Gamma_{r0aug} = 4 / ns. \quad (A33)$$

Due to the low densities of the EHP, the Coulomb interaction between electrons and holes is very weak so that they cool down independently. Therefore, Eq. (A29) has to be solved for electrons and holes separately, leading to the temperatures shown in Fig. A6 by the blue lines. The effective temperature of the EHP was then taken as the arithmetic mean of both temperatures and shows rather good agreement.

Appendix III: Results of the many-body theory

$n = 4$

| $T_{eh} / \rho_{eh}$ | 0.01 | 0.016 | 0.025 | 0.04 | 0.063 | 0.1 | 0.158 | 0.251 | 0.398 | 0.631 | 1.000 |
|---|---|---|---|---|---|---|---|---|---|---|---|
| 1 | -0.002 | -0.002 | -0.002 | -0.002 | -0.002 | -0.002 | -0.002 | -0.002 | -0.002 | -0.002 | -0.002 |
| 2 | -0.003 | -0.003 | -0.004 | -0.004 | -0.004 | -0.004 | -0.004 | -0.004 | -0.004 | -0.004 | -0.004 |
| 4 | -0.004 | -0.005 | -0.006 | -0.006 | -0.006 | -0.006 | -0.006 | -0.006 | -0.006 | -0.006 | -0.006 |
| 6 | -0.007 | -0.008 | -0.009 | -0.009 | -0.009 | -0.01 | -0.01 | -0.01 | -0.01 | -0.01 | -0.01 |
| 8 | -0.011 | -0.013 | -0.014 | -0.015 | -0.015 | -0.015 | -0.015 | -0.015 | -0.016 | -0.016 | -0.016 |
| 10 | -0.017 | -0.02 | -0.022 | -0.023 | -0.024 | -0.024 | -0.024 | -0.024 | -0.025 | -0.025 | -0.025 |
| 12 | -0.027 | -0.031 | -0.035 | -0.036 | -0.037 | -0.038 | -0.038 | -0.039 | -0.039 | -0.039 | -0.039 |
| 14 | -0.043 | -0.049 | -0.055 | -0.058 | -0.059 | -0.06 | -0.061 | -0.061 | -0.062 | -0.062 | -0.062 |
| 16 | -0.068 | -0.078 | -0.087 | -0.091 | -0.093 | -0.095 | -0.096 | -0.097 | -0.098 | -0.098 | -0.099 |
| 18 | -0.106 | -0.123 | -0.137 | -0.144 | -0.148 | -0.15 | -0.152 | -0.154 | -0.155 | -0.156 | -0.156 |
| 20 | -0.167 | -0.193 | -0.216 | -0.227 | -0.233 | -0.238 | -0.241 | -0.243 | -0.245 | -0.246 | -0.248 |

$n = 6$

| $T_{eh} / \rho_{eh}$ | 0.01 | 0.016 | 0.025 | 0.04 | 0.063 | 0.1 | 0.158 | 0.251 | 0.398 | 0.631 | 1.000 |
|---|---|---|---|---|---|---|---|---|---|---|---|
| 1 | -0.013 | -0.014 | -0.014 | -0.015 | -0.015 | -0.015 | -0.015 | -0.015 | -0.015 | -0.015 | -0.015 |
| 2 | -0.021 | -0.022 | -0.023 | -0.023 | -0.023 | -0.023 | -0.023 | -0.024 | -0.024 | -0.024 | -0.024 |
| 4 | -0.034 | -0.035 | -0.036 | -0.036 | -0.037 | -0.037 | -0.037 | -0.037 | -0.038 | -0.038 | -0.038 |
| 6 | -0.053 | -0.056 | -0.057 | -0.058 | -0.058 | -0.059 | -0.059 | -0.059 | -0.06 | -0.06 | -0.06 |
| 8 | -0.084 | -0.088 | -0.09 | -0.091 | -0.092 | -0.093 | -0.093 | -0.094 | -0.094 | -0.095 | -0.095 |
| 10 | -0.133 | -0.14 | -0.143 | -0.144 | -0.146 | -0.147 | -0.148 | -0.148 | -0.149 | -0.15 | -0.15 |
| 12 | -0.21 | -0.221 | -0.225 | -0.228 | -0.23 | -0.232 | -0.234 | -0.235 | -0.236 | -0.238 | -0.238 |
| 14 | -0.331 | -0.349 | -0.356 | -0.361 | -0.364 | -0.367 | -0.37 | -0.372 | -0.374 | -0.376 | -0.376 |
| 16 | -0.521 | -0.55 | -0.563 | -0.57 | -0.576 | -0.58 | -0.584 | -0.588 | -0.592 | -0.595 | -0.595 |
| 18 | -0.819 | -0.867 | -0.888 | -0.9 | -0.909 | -0.917 | -0.924 | -0.93 | -0.936 | -0.941 | -0.941 |
| 20 | -1.285 | -1.365 | -1.4 | -1.42 | -1.435 | -1.448 | -1.459 | -1.469 | -1.478 | -1.487 | -1.487 |

$n = 8$

| $T_{eh}/\rho_{eh}$ | 0.01 | 0.016 | 0.025 | 0.04 | 0.063 | 0.1 | 0.158 | 0.251 | 0.398 | 0.631 | 1.000 |
|---|---|---|---|---|---|---|---|---|---|---|---|
| 1 | -0.046 | -0.048 | -0.049 | -0.049 | -0.05 | -0.05 | -0.051 | -0.051 | -0.051 | -0.052 | -0.052 |
| 2 | -0.072 | -0.075 | -0.077 | -0.078 | -0.079 | -0.08 | -0.08 | -0.081 | -0.081 | -0.082 | -0.082 |
| 4 | -0.114 | -0.119 | -0.122 | -0.124 | -0.125 | -0.126 | -0.127 | -0.128 | -0.129 | -0.129 | -0.13 |
| 6 | -0.179 | -0.188 | -0.193 | -0.196 | -0.198 | -0.2 | -0.201 | -0.203 | -0.204 | -0.205 | -0.206 |
| 8 | -0.283 | -0.296 | -0.306 | -0.31 | -0.313 | -0.316 | -0.319 | -0.321 | -0.323 | -0.324 | -0.326 |
| 10 | -0.444 | -0.467 | -0.482 | -0.49 | -0.495 | -0.5 | -0.504 | -0.507 | -0.511 | -0.513 | -0.516 |
| 12 | -0.698 | -0.735 | -0.761 | -0.773 | -0.782 | -0.79 | -0.797 | -0.802 | -0.808 | -0.812 | -0.816 |
| 14 | -1.093 | -1.156 | -1.2 | -1.22 | -1.235 | -1.248 | -1.258 | -1.268 | -1.276 | -1.284 | -1.29 |
| 16 | -1.706 | -1.814 | -1.888 | -1.923 | -1.948 | -1.969 | -1.987 | -2.002 | -2.016 | -2.029 | -2.039 |
| 18 | -2.654 | -2.838 | -2.966 | -3.026 | -3.068 | -3.103 | -3.133 | -3.159 | -3.182 | -3.202 | -3.219 |
| 20 | -4.111 | -4.427 | -4.648 | -4.753 | -4.826 | -4.885 | -4.935 | -4.978 | -5.016 | -5.049 | -5.078 |

$n = 10$

| $T_{eh}/\rho_{eh}$ | 0.01 | 0.016 | 0.025 | 0.04 | 0.063 | 0.1 | 0.158 | 0.251 | 0.398 | 0.631 | 1.000 |
|---|---|---|---|---|---|---|---|---|---|---|---|
| 1 | -0.123 | -0.125 | -0.127 | -0.128 | -0.129 | -0.13 | -0.131 | -0.131 | -0.132 | -0.132 | -0.123 |
| 2 | -0.194 | -0.198 | -0.201 | -0.203 | -0.205 | -0.206 | -0.207 | -0.208 | -0.209 | -0.21 | -0.194 |
| 4 | -0.306 | -0.313 | -0.318 | -0.321 | -0.324 | -0.326 | -0.328 | -0.329 | -0.331 | -0.332 | -0.306 |
| 6 | -0.483 | -0.495 | -0.502 | -0.508 | -0.512 | -0.516 | -0.519 | -0.521 | -0.523 | -0.525 | -0.483 |
| 8 | -0.761 | -0.78 | -0.793 | -0.802 | -0.809 | -0.816 | -0.821 | -0.825 | -0.828 | -0.831 | -0.761 |
| 10 | -1.196 | -1.231 | -1.251 | -1.266 | -1.279 | -1.289 | -1.297 | -1.304 | -1.309 | -1.314 | -1.196 |
| 12 | -1.878 | -1.938 | -1.972 | -1.998 | -2.018 | -2.035 | -2.048 | -2.059 | -2.068 | -2.076 | -1.878 |
| 14 | -2.94 | -3.046 | -3.104 | -3.148 | -3.182 | -3.21 | -3.232 | -3.25 | -3.265 | -3.279 | -2.94 |
| 16 | -4.59 | -4.776 | -4.878 | -4.952 | -5.01 | -5.057 | -5.094 | -5.125 | -5.15 | -5.172 | -4.59 |
| 18 | -7.138 | -7.47 | -7.648 | -7.776 | -7.876 | -7.955 | -8.019 | -8.072 | -8.114 | -8.151 | -7.138 |
| 20 | -11.05 | -11.65 | -11.96 | -12.18 | -12.36 | -12.49 | -12.6 | -12.69 | -12.77 | -12.83 | -11.05 |

$n = 12$

| $T_{eh}/\rho_{eh}$ | 0.01 | 0.016 | 0.025 | 0.04 | 0.063 | 0.1 | 0.158 | 0.251 | 0.398 | 0.631 | 1.000 |
|---|---|---|---|---|---|---|---|---|---|---|---|
| 1 | -0.257 | -0.263 | -0.268 | -0.272 | -0.274 | -0.276 | -0.278 | -0.279 | -0.28 | -0.281 | -0.281 |
| 2 | -0.404 | -0.415 | -0.424 | -0.43 | -0.434 | -0.437 | -0.44 | -0.442 | -0.443 | -0.444 | -0.445 |
| 4 | -0.635 | -0.654 | -0.669 | -0.679 | -0.686 | -0.691 | -0.696 | -0.699 | -0.701 | -0.703 | -0.704 |
| 6 | -0.997 | -1.029 | -1.055 | -1.072 | -1.084 | -1.093 | -1.099 | -1.105 | -1.108 | -1.112 | -1.114 |
| 8 | -1.561 | -1.617 | -1.663 | -1.691 | -1.711 | -1.725 | -1.737 | -1.745 | -1.752 | -1.757 | -1.761 |
| 10 | -2.437 | -2.535 | -2.615 | -2.663 | -2.697 | -2.722 | -2.741 | -2.755 | -2.766 | -2.775 | -2.783 |
| 12 | -3.79 | -3.964 | -4.106 | -4.189 | -4.246 | -4.289 | -4.321 | -4.345 | -4.364 | -4.379 | -4.392 |
| 14 | -5.865 | -6.178 | -6.431 | -6.576 | -6.675 | -6.747 | -6.802 | -6.844 | -6.877 | -6.903 | -6.924 |
| 16 | -9.025 | -9.588 | -10.04 | -10.30 | -10.47 | -10.60 | -10.69 | -10.76 | -10.82 | -10.87 | -10.90 |
| 18 | -13.79 | -14.81 | -15.63 | -16.08 | -16.38 | -16.60 | -16.77 | -16.90 | -17.00 | -17.08 | -17.14 |
| 20 | -20.89 | -22.72 | -24.21 | -25.01 | -25.55 | -25.94 | -26.23 | -26.46 | -26.63 | -26.78 | -26.90 |

Table A2

The table gives the numbers calculated by the theory of Ref. 42 from which Eq. (9) of the main text was derived by multivariate regression. The density is given in $10^{10} \text{cm}^{-3}$, the temperature in K, the error of the fit formula is less than $0.1 \mu \text{eV}$.

Figures

Fig.1

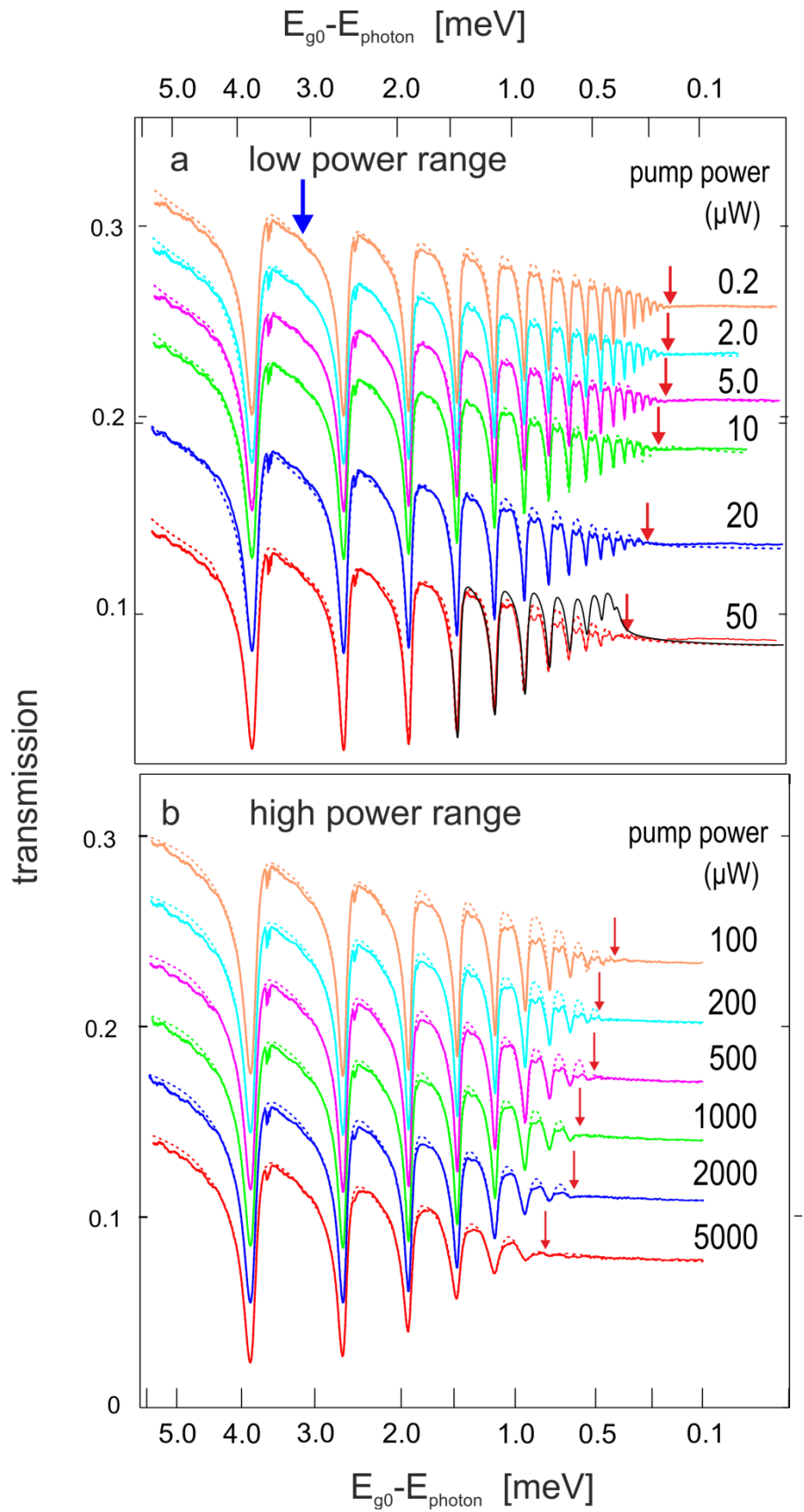

Fig. 1: Transmission spectra of Rydberg excitons from n = 5 onwards recorded at T = 1.35 K for simultaneous application of a pump laser at a photon energy of 2.1688830 eV (blue arrow) operated at different powers as shown. The full lines are the experimental results and the dotted lines show the best fit using the CTM method [46].

The upper panel shows the spectra at low pump powers (LPR) which were fitted assuming an Urbach-like tail (Eq. (6)). The black line for 50 $\mu$W is a fit without an Urbach tail.

The lower panel shows the spectra for high pump powers (HPR), which were fitted assuming only a broadened continuum (Eq. (5)). The deviations of the experimental traces from the fit between the P lines are due to optical absorption into other angular momentum states that become allowed through the effect of the impurity induced electric fields [39].

Note the non-linear scale of the x-axis proportional to the square root of energy difference of nominal band edge $E_{g0}$ and photon energy allowing displaying all lines with equal resolution. The red arrows mark the position of the apparent band edge $E_g$, where the absorption lines of the P states vanish and the transmission goes over to the flat continuum. The band-edge shifts $\Delta = E_g - E_{g0}$ are plotted as black squares in Fig. 4a.

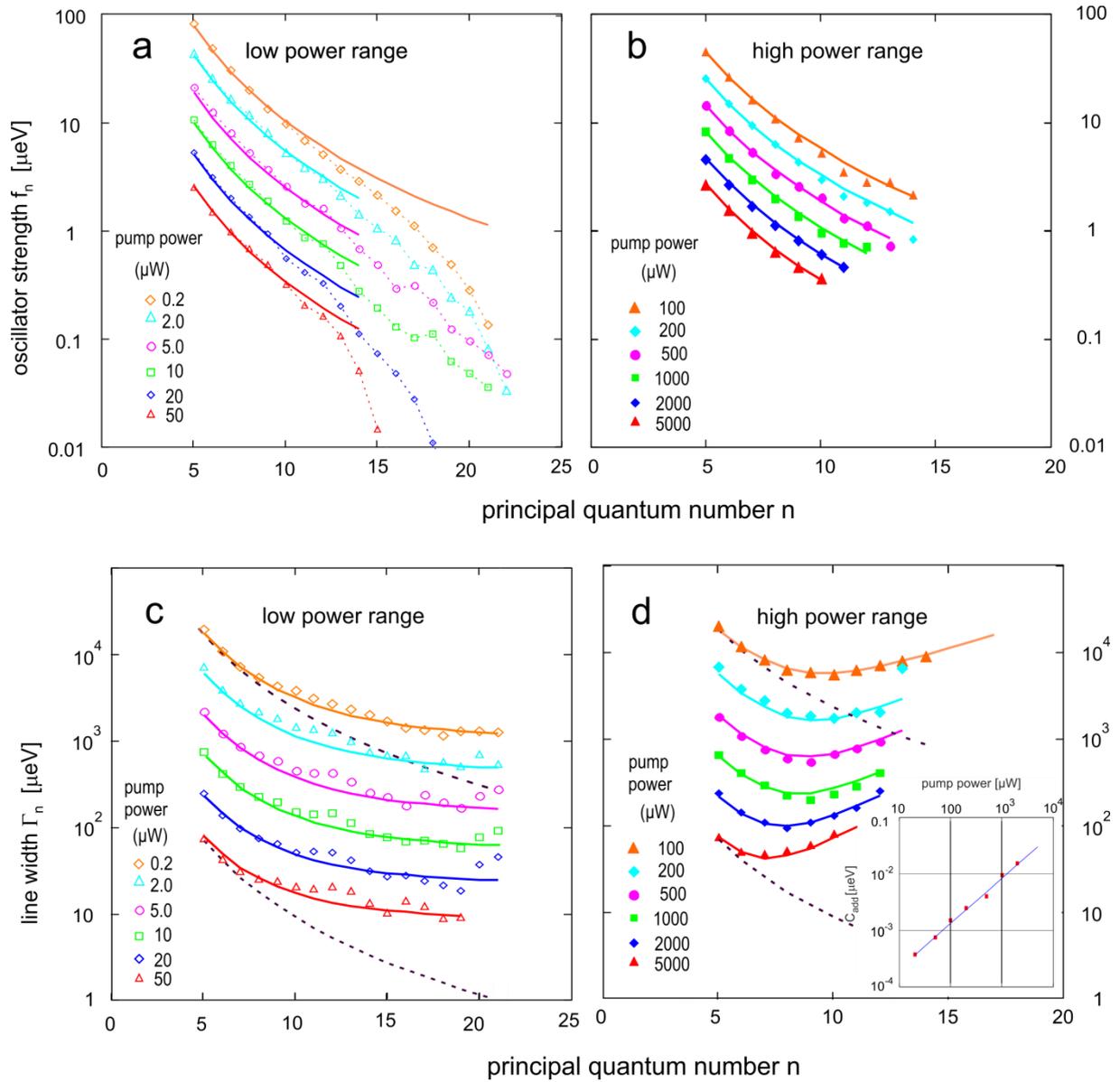

Fig. 2: Results of the fit of the transmission spectra for various pump powers. Panels a and b: oscillator strengths for different pump powers from 0.2 µW to 5 mW grouped into low and high power ranges. The pump powers are the same as in Fig. 1. The colored full lines give the dependence of oscillator strength on quantum number according to exciton theory as discussed in the text. While in the HPR (panel b) the fit results (within experimental error) all fall onto this line, in the LPR (panel a) we see the well-known deviations due to the effect of charged impurities. Panels c and d show the dependence of line width of the Lorentzians (given as HWHM $\Gamma_n$) on principal quantum number n. The different pump powers are represented by the same symbols as in a and b. The dashed black lines give the expected dependence from phonon scattering, while the colored lines contain an additional contribution as discussed in the text. The inset in panel d shows the pump-power dependence of the strength factor $C_{add}$. The blue full line gives a pump-power dependence $P^{0.8}$.

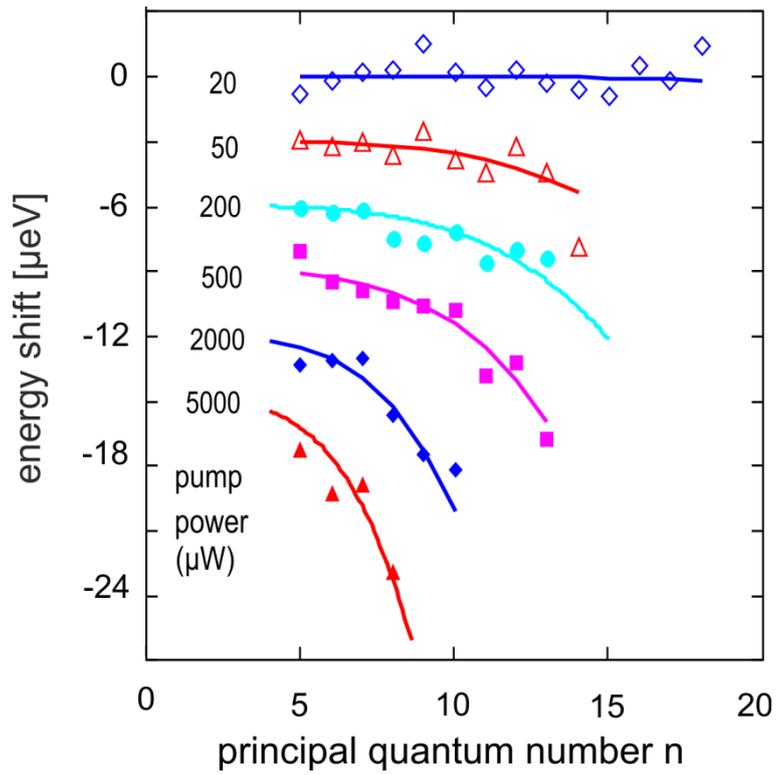

Fig. 3:
Dependence of energy shifts on the principal quantum number n for selected pump powers from 20 µW to 5 mW as indicated. The full lines give the results of many-particle theory proportional to the fourth power of the principal quantum number (Eq. (9)). The proportionality constants allow deducing the plasma densities and temperatures shown in Fig. 4b, c as blue triangles.

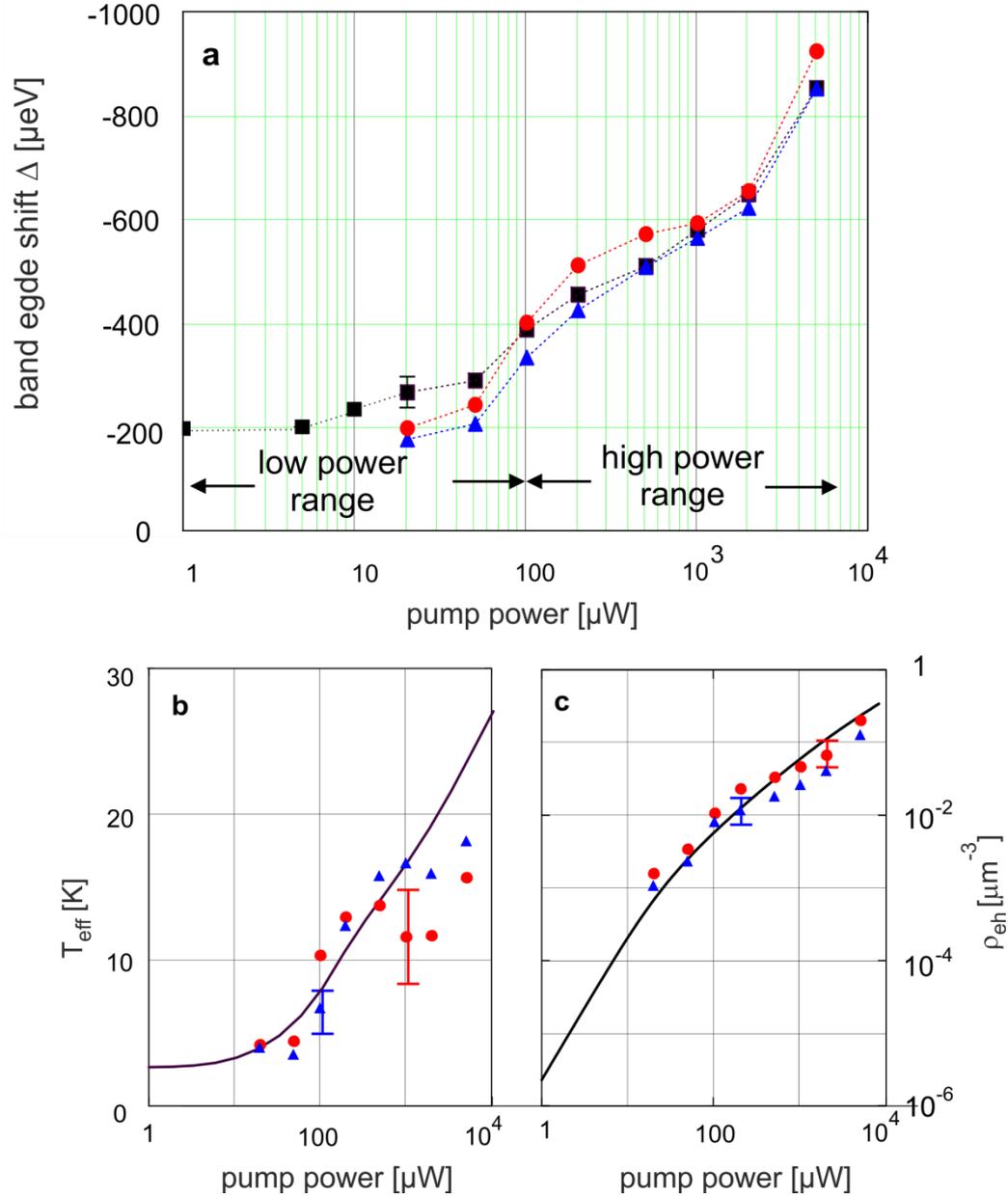

Fig. 4: Panel a shows the shift of the band edge $\Delta$ as a function of the pump power. Black square symbols are the shifts as determined from the transmission spectra (red arrows in Fig. 1), while the red dots and the blue triangles are the shifts calculated by many-body theory (Eq. (1.4)) from the densities and temperatures obtained from fitting the linewidth broadening (Fig. 2b) and the shift of the resonance energies (Fig. 3), respectively. In the calculation we neglected the contribution of the charged impurities, resulting in a too small shift at low pump powers, as expected.

Panel b gives the effective temperatures of the EHP obtained from the line-energy and band-edge shifts (blue triangles) and from the line broadening (red dots) compared to results of the relaxation and cooling model (see Appendix II) for the phonon-assisted Auger scenario (black line).

Panel c shows the dependence of the electron-hole pair concentration on pump power obtained from the P lines energy shift and the band-edge shift (blue triangles), from the linewidth broadening (red dots) and from the rate and cooling model (black line).

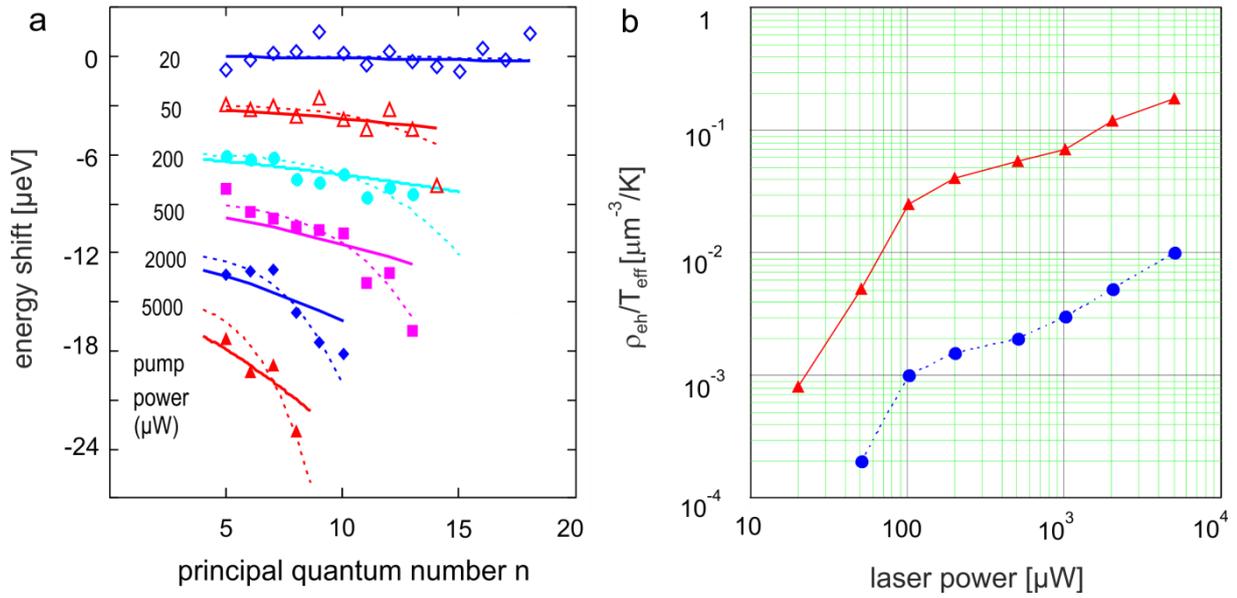

Fig. 5: Breakdown of the Debye plasma model: Panel a compares the variations of the line-energy shifts on principal quantum number n for selected pump powers from 20 to 5000 μW as indicated (same data as in Fig. 3) with the predictions of the Debye plasma model (Eq. (10), full lines). The dotted lines give the results of the many-particle theory (Eq. (9)) as taken from Fig. 3. The ratios of density and temperature obtained from the proportionality constants are plotted as blue dots in panel b, while the red triangles show the same quantity obtained from the band-edge shift (Eq. (11)).

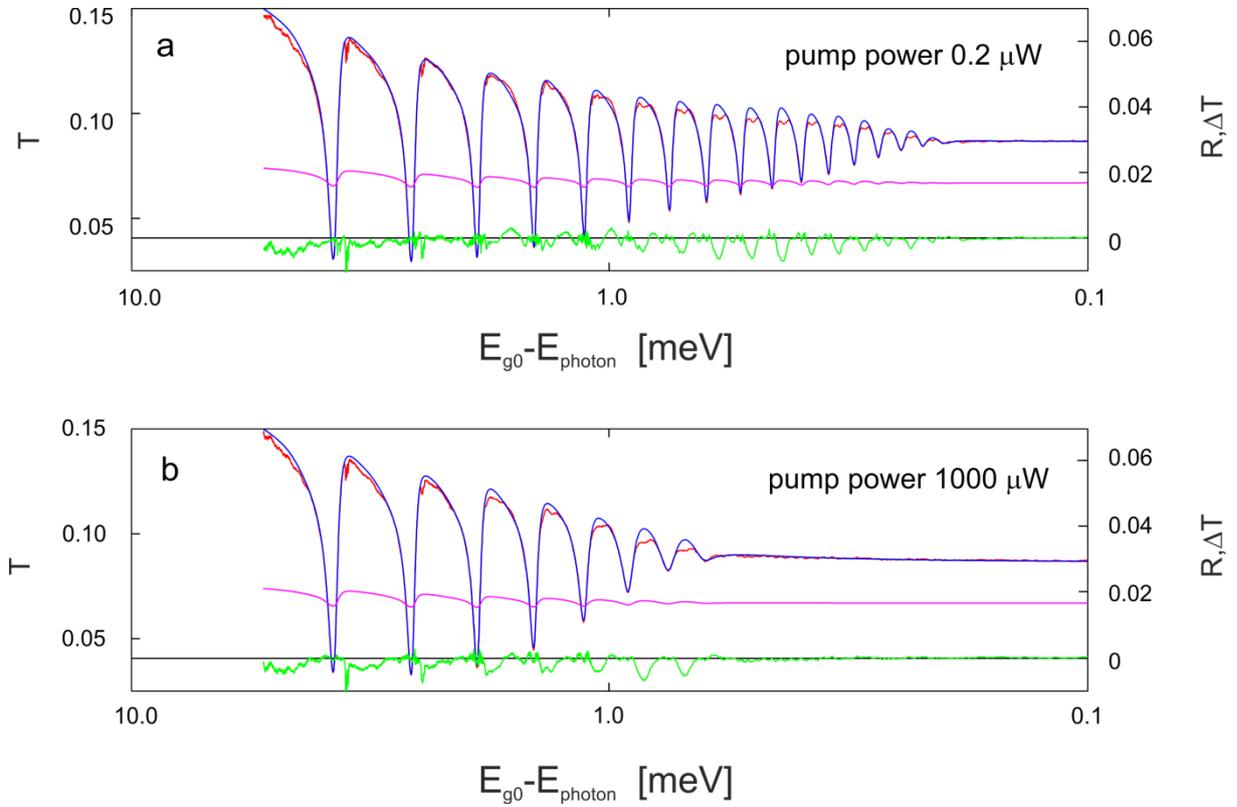

Fig. A1: Comparison of the measured transmission spectrum (red) with the fit according to CTM theory (blue). For low (panel a) and high (panel b) pump power as indicated. The difference between both transmission spectra is shown as the green line. While one clearly sees residual absorption lines into F states (from $n=5$ to $7$) [60], the deviations at higher principal quantum numbers ($n>10$) are due to states that become allowed by the electric field of residual charged impurities [39] and to interference effects of the setup [46]. The magenta line gives the corresponding theoretical reflection spectrum (shifted by 0.15). Pump laser: photon energy 2.168883 eV.

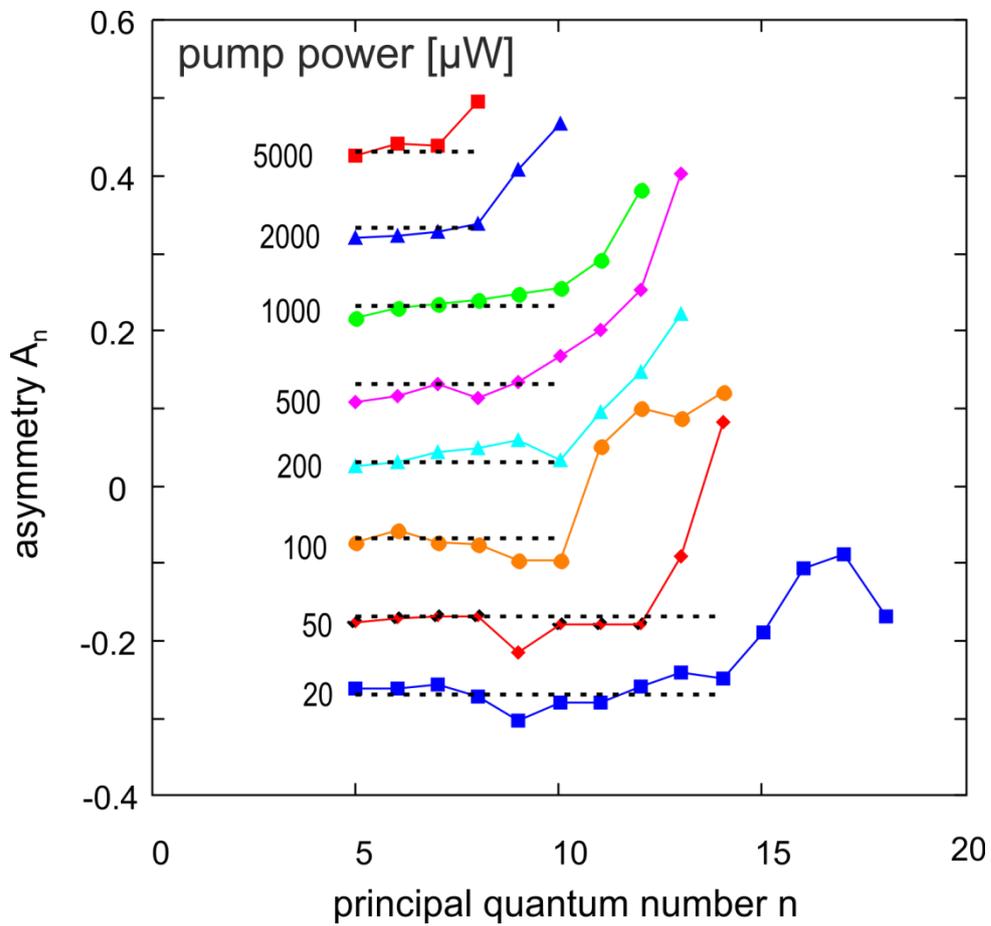

Fig. A2:
Overview of the asymmetry parameters from the fits of the transmission spectra. The pump powers are given at the left of the curves. The full lines are guides to the eye, the dotted lines gives the constant value of $A = -0.275$ at low principal quantum numbers for all measurements. For increasing pump power the lines are shifted by 0.1 from data set to data set.

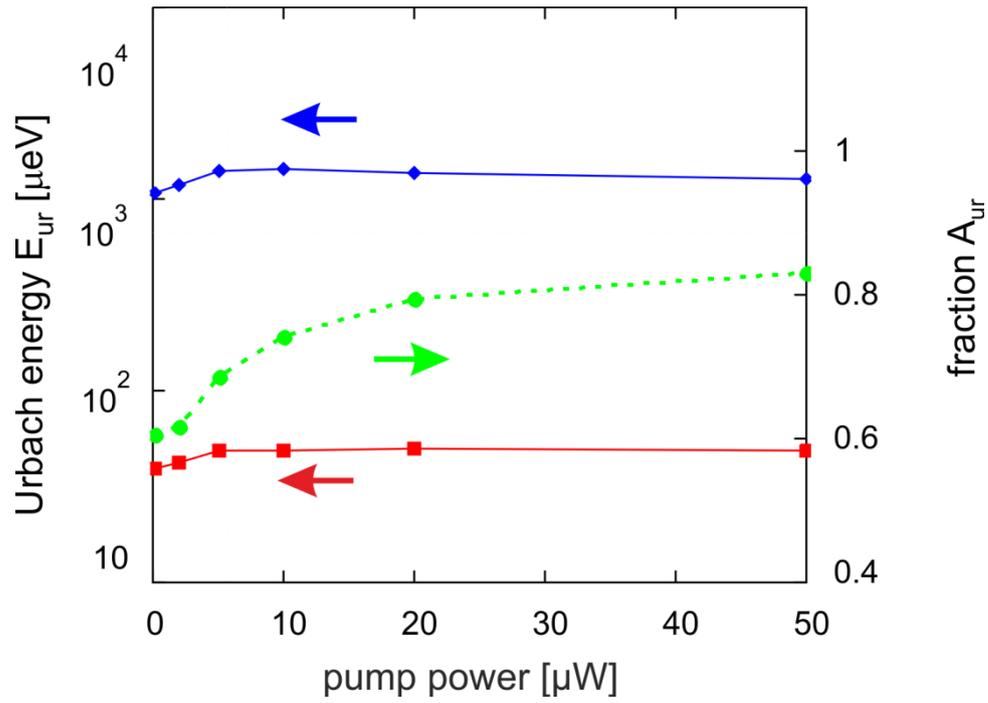

Fig. A3: Urbach tail decay parameters from Eq. (A2). $E_{ur1}$ (red squares), $E_{ur2}$ (blue diamonds), and $A_{ur}$ (green dots) on pump power in the LPR. The lines are guides to the eye.

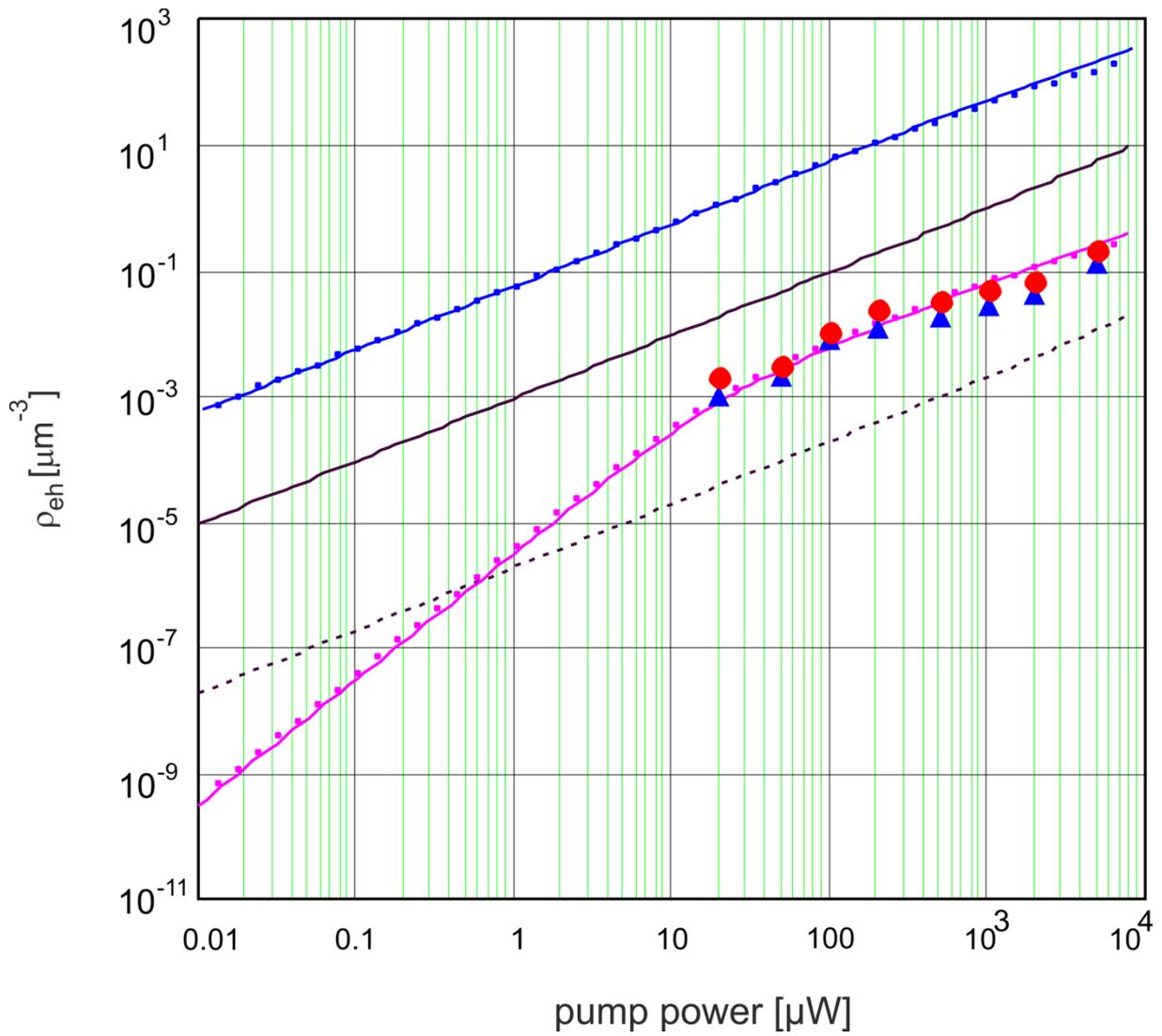

Fig. A4: Results of the rate model. The curves give the concentrations of the following species: dashed black line: primarily excited orthoexcitons, full black line: orthoexcitons at k=0, full blue line: paraexcitons at k=0, magenta line: Auger created eh pairs, the dotted lines of the same color give the results of the effective rate model used to simulate the cooling behavior (see Section A IIb). The red dots and blue triangles give the results of the analysis of the experimental data (see main text).

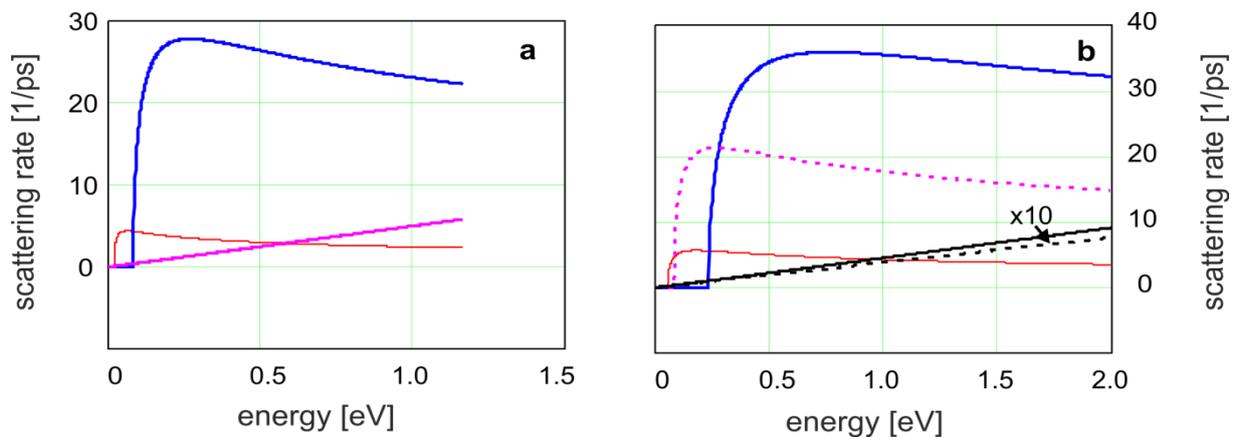

Fig. A5: Scattering rates of electrons (a) and holes (b) versus kinetic energy. Blue lines denote scattering by the LO2 mode via Fröhlich interaction, red lines that of the LO1 mode. In panel a scattering by acoustic phonons is denoted by the magenta line. In panel b scattering by the LO2 mode in the lower part of the valence band is denoted by a dashed magenta line, while scattering by acoustical phonons by a black line. The dashed black line gives the acoustical scattering in the upper part of the valence band (note the multiplication by a factor of 10).

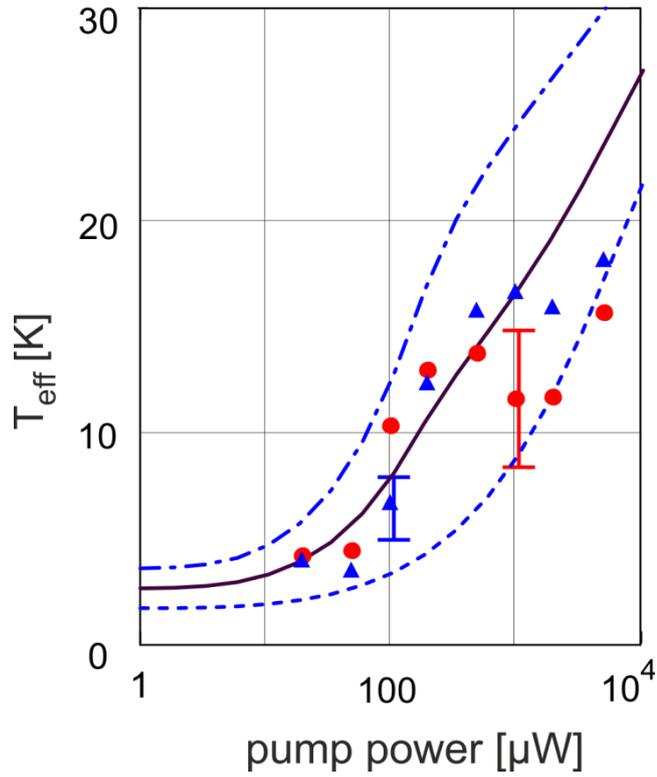

Fig. A6: Temperatures of the EHP versus excitation laser power. The blue triangles and red dots denote the values derived from the analysis of the line shifts and line broadening, respectively (see main text). The black line gives the result of the calculation for the average temperature if we assume that the phonon-assisted Auger scattering is dominant (see text). The blue dashed line gives the temperature of the electrons alone, while the dash-dotted blue line gives the temperature of the holes.